\documentclass{aa}

\usepackage{graphicx}
\usepackage{mathrsfs}  
\usepackage{natbib}
\usepackage{amssymb}
\bibpunct{(}{)}{;}{a}{}{,}
\usepackage{caption}
\usepackage[varg]{txfonts}
\usepackage[section]{placeins}
\usepackage[english]{babel}
\usepackage{url}
\usepackage{arydshln}
\usepackage{color}

\def\phn{\phantom{0}}

\authorrunning{Magee et al.}
\titlerunning{SN~2015H}

\begin{document} 
            
    \title{The type Iax supernova, SN~2015H: }
    \subtitle{a white dwarf deflagration candidate }
	\author{M. R. Magee\inst{1}{\thanks{E-mail: mmagee37@qub.ac.uk}}
    		\and R. Kotak\inst{1}
            \and S. A. Sim\inst{1}
            \and M. Kromer\inst{2}
            \and D. Rabinowitz\inst{3}
            \and S. J. Smartt\inst{1}
            \and C. Baltay\inst{3}
            \and H. C. Campbell\inst{4}              			
            \and T.-W. Chen\inst{5}
            \and M. Fink\inst{6}
      		\and A. Gal-Yam\inst{7}
            \and L. Galbany\inst{8,9}
            \and W. Hillebrandt\inst{10}
            \and C. Inserra\inst{1}
            \and E. Kankare\inst{1}
            \and L. Le Guillou\inst{11,12}
            \and J. D. Lyman\inst{13}
            \and K. Maguire\inst{1}
            \and R. Pakmor\inst{14}
            \and F. K. R{\"o}pke\inst{14}
            \and A. J. Ruiter\inst{15,16}
            \and I. R. Seitenzahl\inst{15,16}
            \and M. Sullivan\inst{17}
            \and S. Valenti\inst{18}
            \and D. R. Young\inst{1}
            }

	\institute{Astrophysics Research Centre, School of Mathematics and Physics, Queen's University 			Belfast, Belfast, BT7 1NN, UK
    \and The Oskar Klein Centre \& Department of Astronomy, Stockholm University, AlbaNova, SE-106 91 			Stockholm Sweden
   \and Department of Physics, Yale University, New Haven, CT 06250-8121, USA 
   \and Institute of Astronomy, University of Cambridge, Madingley Road, Cambridge
		CB3 0HA, UK 
   \and Max-Planck-Institut f{\"u}r Extraterrestrische Physik, Giessenbachstra\ss e 1, 85748, 				Garching, Germany
   \and Institut f{\"u}r Theoretische Physik und Astrophysik, Universit{\"a}t W{\"u}rzburg, Emil-			Fischer Stra\ss e 31,97074 W{\"u}rzburg, Germany
    \and Department of Particle Physics and Astrophysics, Weizmann Institute of Science, Rehovot 			76100, Israel
    \and Milenium Institute of Astrophysics, Santiago, Chile
    \and Departamento de Astronom{\'i}a, Universidad de Chile, Camino El Observatorio 1515, Las 			Condes, Santiago, Chile
    \and Max-Planck-Institut f{\"u}r Astrophysik, Karl-Schwarzschild-Str. 1, D-85748 Garching bei 			M{\"u}nchen, Germany
    \and Sorbonne Universites, UPMC Univ. Paris 06, UMR 7585, LPNHE, F-75005
Paris, France
	\and CNRS, UMR 7585, Laboratoire de Physique Nucleaire et des Hautes
Energies, 4 place Jussieu, 75005 Paris, France 
    \and Department of Physics, University of Warwick, Coventry CV4 7AL, UK
	\and Heidelberger Institut f{\"u}r Theoretische Studien, Schloss-Wolfsbrunnenweg 35, 69118 Heidelberg, Germany
    \and Research School of Astronomy and Astrophysics, Australian National University, Canberra, ACT 2611, Australia
    \and ARC Centre of Excellence for All-Sky Astrophysics (CAASTRO)
    \and Department of Physics and Astronomy, University of Southampton, Southampton, SO17 1BJ, UK
    \and Department of Physics, University of California, Davis, CA 95616, USA
		}

   \date{Received -	- -; accepted - - - }


 \abstract{We present results based on observations of SN~2015H which belongs to the small group of objects similar to SN~2002cx, otherwise known as type Iax supernovae. The availability of deep pre-explosion imaging allowed us to place tight constraints on the explosion epoch. Our observational campaign began approximately one day post-explosion, and extended over a period of about 150 days post maximum light, making it one of the best observed objects of this class to date. We find a peak magnitude of $M_r = -17.27\pm$0.07, and a $(\Delta m_{15})_{r}$ = 0.69$\pm$0.04. Comparing our observations to synthetic spectra generated from simulations of deflagrations of Chandrasekhar mass carbon-oxygen white dwarfs, we find reasonable agreement with models of weak deflagrations that result in the ejection of $\sim$0.2\,M$_\odot$ of material containing $\sim$0.07 M$_\odot$ of $^{56}$Ni. The model light curve however, evolves more rapidly than observations, suggesting that a higher ejecta mass is to be favoured. Nevertheless, empirical modelling of the pseudo-bolometric light curve suggests that $\lesssim$0.6\,M$_\odot$ of material was ejected, implying that the white dwarf is not completely disrupted, and that a bound remnant is a likely outcome.}


   \maketitle
%

\section{Introduction}
\label{sect:intro}
The use of Type Ia supernovae (SNe Ia) as standardizable candles and cosmological distance indicators \citep{perlmutter99, riess--expansion} has meant that they are among the best studied transient phenomena in the  Universe. In spite of this however, identifying the  nature of the underlying progenitor system(s) remains a key challenge. While it is generally agreed that SNe Ia result from the thermonuclear explosion of a carbon-oxygen white dwarf (CO WD), the explosion mechanism and scenarios leading to explosion remain unclear \citep[e.g.][]{hillebrandt--2013}. During the course of transient surveys, some of which are dedicated to searching for SNe Ia, many peculiar transients have been discovered. A small subset of these show similarities to SNe Ia, but also striking differences. If these also result from thermonuclear explosions, then they may prove to be an excellent opportunity to test the extreme boundaries of explosion models.  

\par
One such class of peculiar SNe Ia are the SN~2002cx-like objects, that have recently been dubbed `SNe Iax' \citep{02cx--orig, foley--13}. 
The first examples of this group included SN~2002cx which was immediately recognized as being different from SNe Ia \citep{02cx--orig}. A few years later, it was followed by SN~2005hk \citep{05hk--deflag?}. Both objects showed similarities to each other, but diverged from usual SNe Ia behaviour. A subsequent search yielded $\sim$25 objects that were deemed to exhibit properties in common with SNe~2002cx and 2005hk \citep{foley--13}. 
This sample contains some very well observed SNe Iax such as SN~2012Z, for which a pre-explosion point source coincident with the supernova \citep{12z--prog}.
However, it also necessarily included some objects with incomplete data sets.
The main observational properties that link SNe Iax include slow expansion velocities -- roughly half that of normal SNe Ia at similar epochs, -- spectra that are dominated by intermediate mass (IME) and iron-group elements (IGE) \citep{02cx--orig, read--02cx--spectra}, and peak absolute brightnesses that span about five magnitudes:
 \citep[$-14\gtrsim M_V \gtrsim -19$;][]{02cx--orig, read--02cx--spectra,obs--08ha, foley--13}.

\par
Despite the fact that SNe Iax are generally sub-luminous compared to normal\footnote{We use the term ``normal'' to indicate ``Branch-normal'' type Ia SNe \citep{branch--normal}} SNe Ia, their early-time spectra (i.e., before or around maximum light), are intriguingly similar to those of over-luminous SNe Ia such as SN~1991T \citep{91t--disc}. Indeed, the prototypical member of the SN Iax class, SN~2002cx, exhibited  features of \ion{Fe}{iii} in pre-maximum spectra \citep{02cx--orig}, similar to SN~1991T-like objects \citep{91t--like}. Spectral features due to \ion{Si}{ii}, \ion{S}{ii}, and \ion{Ca}{ii}, that are prominent in the spectra of normal SNe Ia, have been shown to be present in SNe Iax \citep{read--02cx--spectra}. However, it is at later epochs still, that the differences between SNe Ia and the SNe Iax become most pronounced. While spectra of the former are dominated by forbidden emission lines of IGEs, spectra of the latter exhibit a forest of permitted lines from IGEs. This may indicate that the ejecta density is higher compared to SNe Ia at comparable epochs \citep{05hk--400days}.           

\par
Given that SNe Iax differ markedly in their properties when compared to SNe Ia, alternative explosion mechanisms, and/or progenitor systems are probably necessary. A possible explosion model for SNe Iax that has been widely considered since the early discoveries is that of pure deflagrations of white dwarfs \citep{read--02cx--spectra, 05hk--deflag?}. These models  differ from contemporary explosion models  proposed for normal SNe Ia in that they do not involve a detonation phase \citep{ia--exp--models}. Deflagration models could naturally explain some of the observed aspects of SNe Iax such as their lower kinetic energy and lower luminosity compared to normal SNe Ia. 

\par
Indeed, synthetic observables obtained from deflagration models presented by \cite{3d--deflag--sim--obs}, \cite{3d--deflag--sim--rem, cone--deflag--sim}, and \cite{pure--deflag--3dsim} have shown broad agreement with SNe Iax observables at early times. Nevertheless, the models diverge from observations at later times, indicating that we are missing key pieces in the puzzle that are the SNe Iax. Alternative models to pure deflagrations have also been proposed for SNe Iax; in particular \cite{comp--obs--12z} have shown good agreement for a pulsational delayed detonation model with a bright SN Iax, SN~2012Z. In the context of such models, the range in parameter space spanned by the observations may be attributed to variations in a combination of the details of the ignition as well as inherent diversity in the progenitor systems. It nevertheless remains to be seen whether a single class of model is able to account for the full brightness range spanned by SNe Iax.

\par
We also note in passing that non-thermonuclear explosion scenarios have also been invoked for SNe Iax, in particular, for SN~2008ha. The apparent similarities between SN~2008ha, and some faint core-collapse (II-Plateau) SNe has led to suggestions of a core-collapse origin \citep{low--ccsn,fallback}, usually attributed to stars more massive than $\sim$8\,M$_\odot$. Nevertheless, the undisputed presence of spectral lines due to hydrogen in these type II-Plateau SNe, and the conspicuous absence of these in SNe Iax, including SN~2008ha, poses problems for this scenario. Recent studies that have considered the location of SNe Iax within their host galaxies have shown that SNe Iax tend to favour late-type galaxies with higher local star formation rates, which could indicate that they result from younger stars \citep{low--lum--ia--enviro}. Some progenitor channels involving WDs in binary systems are, however, consistent with the relatively young ages inferred for SNe Iax progenitors \citep{cone--wd+he--prog,cone--deflag--sim} implying that environmental factors alone cannot be used as a discriminating factor. In what follows, we do not consider the core-collapse scenario further, but focus our discussion around thermonuclear explosions, specifically that of a pure delfagration of a CO white dwarf. 

\begin{table*}
\centering
\caption[]{Summary of details for  SN~2015H}
\label{tab:15h--sum}
\begin{tabular}{lll}\hline
\hline
SN~2015H &  & Reference \\
\hline
\hline
Alternative names						& PSN J10544216-2104138 					& \\
 	   												& LSQ15mv			   						& \\
$\alpha$ (J2000.0) 									& $10^{h}54^{m}42^{s}.16$ 								& 1 \\
$\delta$ (J2000.0)									& $-21\degr04'13\farcs7$						& 1 \\
Host galaxy 										& NGC~3464 									& 1 \\
Redshift 											& 0.012462$\pm$0.00001 						& 2 \\
Distance (Mpc)							& 60.57$\pm$1.95					& 2 \\
Distance modulus (mag)        					& 33.91$\pm$0.07 							& 2 \\
Host offset 										& 30$''$E, 14$''$ S 								& 1 \\
Discoverer 											& Backyard Observatory Supernova Search 	& 1 \\
Discovery date (UT) 								& 2015 February 10.545 						& 1 \\
Discovery date (MJD) 								& 57064.05 									& 1 \\
Discovery magnitude (mag)							& +16.9 (Unfiltered) 						& 1 \\
Galactic extinction 								& $E(B-V)$ = 0.048 							& 2 \\
Host extinction 									& $E(B-V)$ = 0 								& 3 \\
$r$--band maximum date (MJD)							& 57061.9$\pm$0.4							& 3 \\
$r$--band maximum magnitude 	(mag)					& 16.77$\pm$0.03 							& 3 \\
$r$--band maximum absolute magnitude 	(mag)		& --17.27$\pm$0.07							& 3 \\
$(\Delta m_{15})_r$ (mag)							& 0.69$\pm$0.04 							& 3 \\
Explosion date (MJD) 								& 57046$\pm$0.5 							& 3 \\
$r$--band rise time 	(days)				            & 15.9$\pm$0.6 								& 3 \\
\hline
\end{tabular}
\tablefoot{(1)~\cite{15h--cbet}, (2) NASA/IPAC Extragalactic Database (NED), (3) This paper }
\end{table*}
\vspace{1cm}
\par
In what follows, we begin by presenting our observations of SN~2015H.
In \S\ref{sect:photometric} and \S\ref{sect:spect-analysis}, we present our analysis of the photometric and spectroscopic evolution, respectively. We compare our data with two well-observed SNe Iax in \S\ref{sect:comp},
and to deflagration models involving Chandrasekhar mass CO WDs in \S\ref{sect:finkcomp}. We summarize our findings in \S\ref{sect:sum}.

\subsection{SN~2015H}

\begin{figure}[!t]
\centering
\includegraphics[width=\columnwidth]{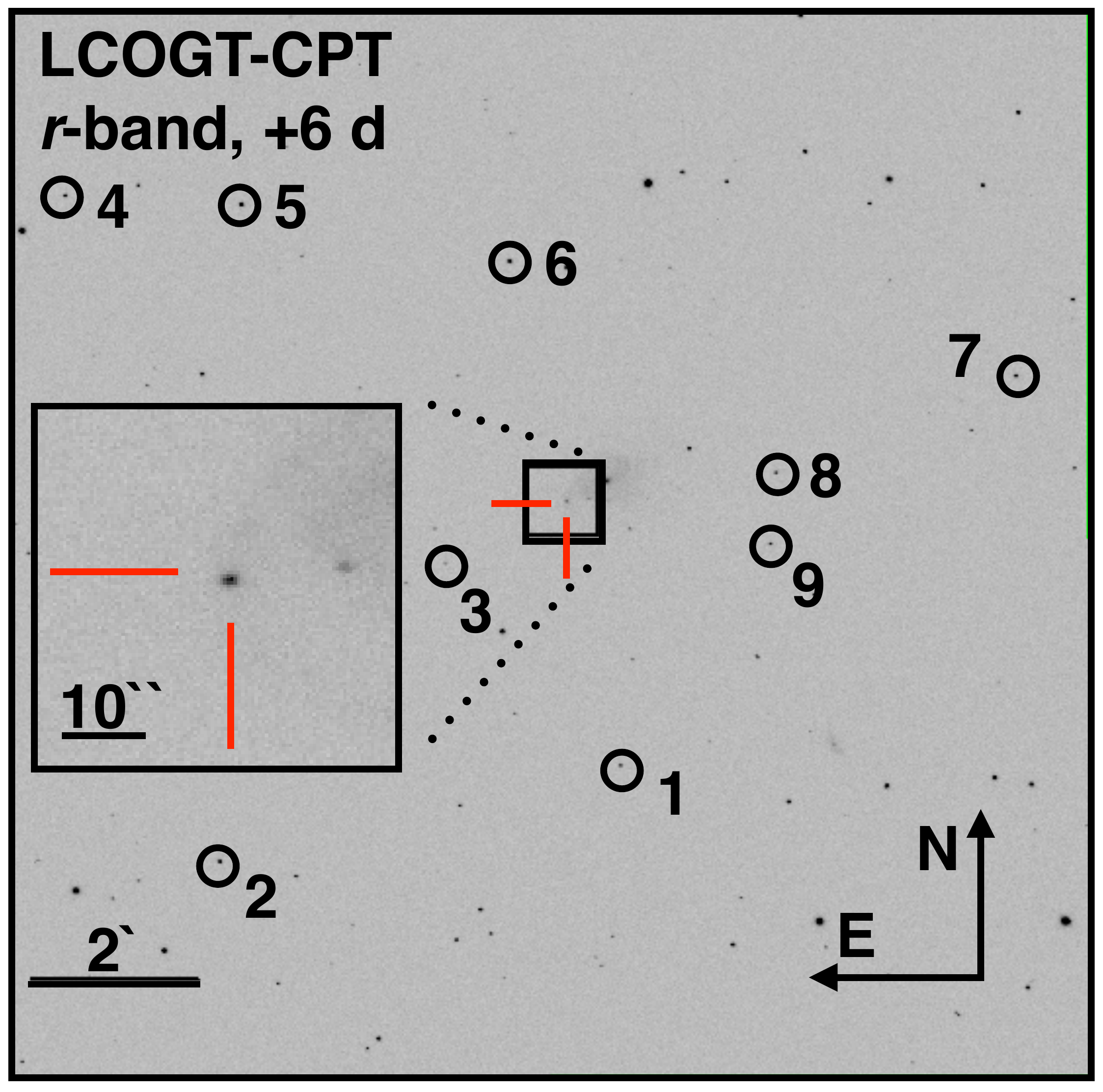}
\caption{$r$--band image of the site of SN~2015H taken approximately one week post $r$--band maximum with the LCOGT (Table 
\ref{tab:15h--phot--opt}). The position of SN~2015H is indicated by red dashes, and lies approximately 10\,kpc from the nucleus of its host galaxy, NGC~3464. The sequence stars marked above were used to calibrate the LCOGT imaging.
\label{15h--field}}
\centering
\end{figure}

One of the most recent additions to the SN Iax class is SN~2015H \citep[also known as LSQ15mv;][]{15h--cbet}. SN~2015H was discovered in an unfiltered CCD image as part of the Backyard Observatory Supernova Search\footnote{http://www.bosssupernova.com/} (BOSS), on 2015 Feb. 10 with an apparent magnitude of +16.9. In Fig.~\ref{15h--field} we show an image of SN~2015H within its host galaxy, NGC~3464, along with sequence stars used to calibrate LCOGT photometry. A classification spectrum obtained by the Public ESO Spectroscopic Survey of Transient Objects \cite[PESSTO,][]{pessto} on 2015 Feb. 11 showed SN~2015H to be a Type Iax SN approximately a few days post-maximum light, with good matches provided by SN~2005cc at +5\,d and SN~2002cx at +10\,d \citep{15h--class}. In what follows, we adopt the distance modulus for NGC~3464 provided by NED\footnote{http://ned.ipac.caltech.edu} of 33.91$\pm$0.07, which is based on the Tully-Fisher relation \citep{tully--fisher}, and derived by \cite{15h--distmod}. In Table~\ref{tab:15h--sum} we list the basic characteristics of SN~2015H and its host galaxy, and our derived parameters for SN~2015H resulting from our analysis in \S~\ref{sect:analysis}.

%

%

\section{Observations \& Data Reduction}
\label{sect:obsandreduc}
\subsection{Optical \& near-IR imaging}
\label{sect:imaging}

A follow-up monitoring campaign for SN~2015H was initiated immediately following the discovery by BOSS and subsequent classification by PESSTO. Although first reported by BOSS, SN~2015H had been serendipitously observed by the La Silla-Quest Variability survey \cite[LSQ,][]{lsq} with images of the field containing SN~2015H extending to approximately three years prior to this date. Incorporating these LSQ images into our analysis of SN~2015H, our full light curve across all filters extends from shortly after explosion (see \S\ref{sect:photometric} for discussion on the explosion date of SN~2015H) to approximately 160\,d later, making it one of the best sampled light curves to date for SNe Iax.  
\par

Optical imaging in the $gri$ filters was provided by the Las Cumbres Observatory Global Telescope 1-m network \cite[LCOGT,][]{lcogt}. We also include the seven $V$--band acquisition images obtained with the 3.58-m New Technology Telescope (NTT) + EFOSC2, as part of the PESSTO spectroscopic monitoring campaign for SN~2015H. The LSQ survey makes use of the 40-inch ESO Schmidt telescope and a `wide' filter approximately covering both the SDSS $g$-- and $r$--bands. This filter function is shown in Fig.~\ref{fig:15h--speccomp}(e), along with standard filters. Three epochs of dithered near-IR $J$-- and $H$--band imaging were obtained with NTT+SOFI, ranging from approximately one week post-maximum to over three weeks later, while time constraints limited us to a single $K_{\mathrm{s}}$--band observation at one week post maximum. 
\par
LSQ and LCOGT images are automatically reduced using their respective pipelines \citep{lsq, lcogt}. NTT images were reduced using the custom-built PESSTO pipeline \cite{pessto}. All of these included standard reduction procedures.

\par
\begin{figure}
\includegraphics[width=\columnwidth]{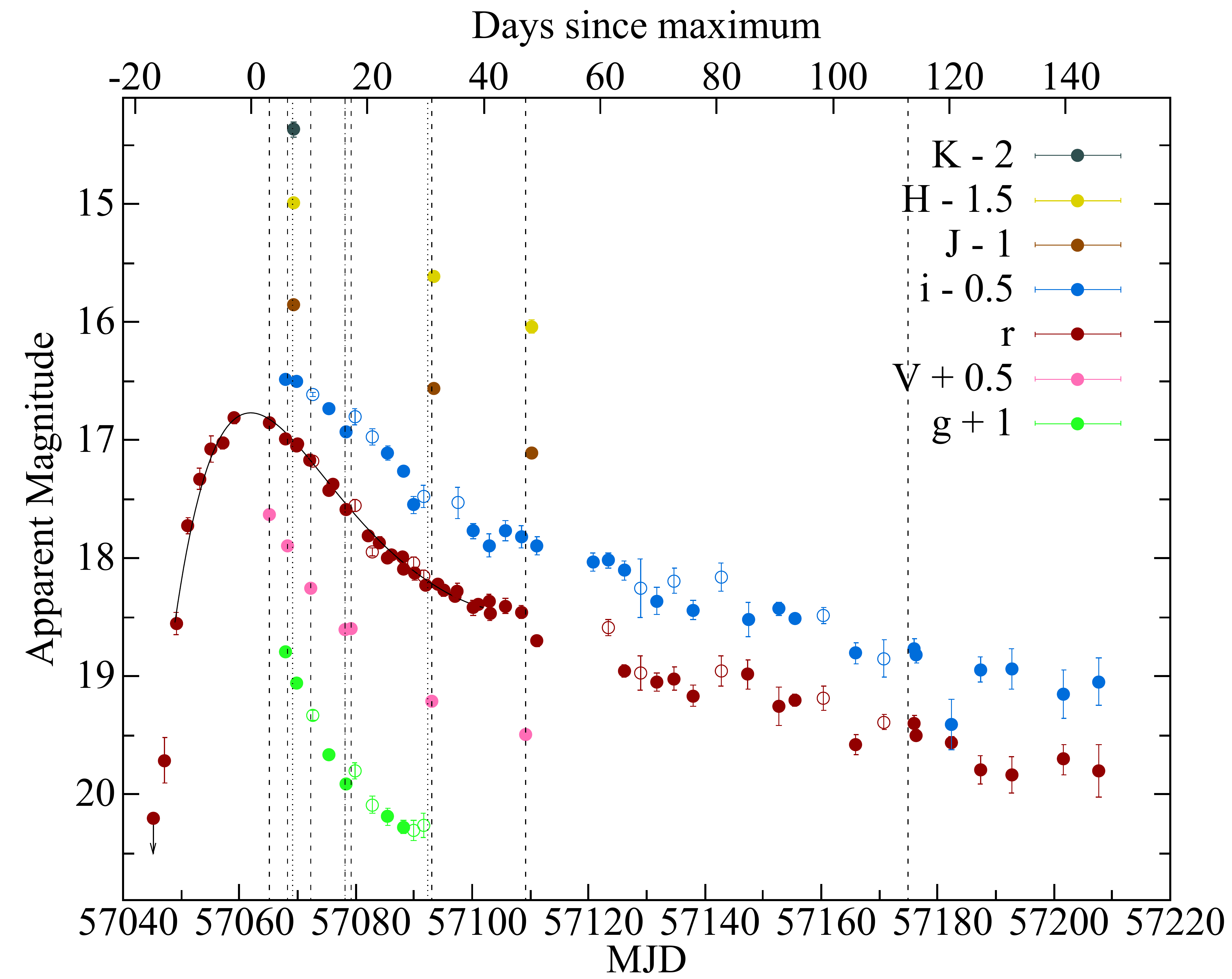}
\caption{Light curve of SN~2015H. We estimate maximum light in the $r$--band to have occurred at MJD = 57061.9. LCOGT images with seeing >2$\arcsec$ are marked with unfilled circles. Note that a vertical offset (indicated) has been applied to each filter. Also shown is a low-order polynomial fit to the LSQ points from which we derive the peak magnitude and epoch of maximum. A pre-explosion limit (with the appropriate vertical offset) from observations taken approximately 48 hours before our first LSQ detection is shown as a filled circle. Vertical dashed lines indicate epochs of optical spectroscopic observations, while dotted lines indicate epochs of near-IR spectroscopy. 
}
\label{fig:15h--lc}
\end{figure}

Photometry on all images was performed using SNOoPY\footnote{http://sngroup.oapd.inaf.it/snoopy.html}, a custom IRAF package designed for point spread function (PSF) fitting photometry. For each image, a collection of stars in the field is used to build up a reference PSF.
We began by subtracting the sky background. The reference PSF was then used to subtract the SN signal, and a new background estimate was derived from the residual image. This process was then repeated in an iterative fashion to finally obtain the SN magnitude. 
Given the availability of pre-explosion imaging from LSQ, we opted to measure the SN brightness from template-subtracted images. For all imaging data, zero points were calculated by calibrating the instrumental magnitudes of the PSF stars to either SDSS magnitudes for optical images, or 2MASS magnitudes for near-IR images. $V$--band magnitudes of the sequence stars were obtained via transformations from SDSS $gri$ \citep{jester--colours}. We transformed the LSQ photometry directly into the SDSS $r$--band by calculating zero points to SDSS $r$--band as a function of $g-r$ colour. 
The differing fields-of-view of the instruments necessitated the use of different sets of sequence stars to calibrate the photometry.
These are shown in Fig. \ref{15h--field} for LCOGT, and listed in Tables \ref{tab:15h--lcogt-std}, \ref{tab:15h--lsq-std}, and \ref{tab:15h--ntt-std} for LCOGT, LSQ, and NTT, respectively. Uncertainties in the SN magnitude are estimated by recovering the magnitudes of artificial stars placed in the region surrounding the SN. In Tables~\ref{tab:15h--phot--opt} and \ref{tab:15h--phot--ir} we present the full set of optical and IR photometric observations obtained for SN~2015H, respectively. Fig.~\ref{fig:15h--lc} shows our complete light curve.

%

 \subsection{Optical \& near-IR spectroscopy}
 \label{sect:spectroscopy}

We present a log of our instrumental configurations and spectroscopic observations in Table~\ref{tab:15h--specseq}. We obtained seven optical spectra using NTT + EFOSC2 and three near-IR spectra using NTT + SOFI. The spectra range from a few days post-maximum to approximately 50 days post-maximum light. We were also able to obtain an optical spectrum using the Gran Telescopio Canarias (GTC) + OSIRIS at approximately 113 days post-maximum light.
\par
All NTT spectra were reduced in a standard fashion using the PESSTO pipeline \citep{pessto}. For spectroscopic data, this includes standard procedures for flux and wavelength calibration. Flux calibration is performed against a spectroscopic standard star observed on the same night, and with the same instrumental set-up. Our GTC spectrum was taken with Grism R500R and reduced using standard IRAF routines. Additionally, we estimated synthetic magnitudes from our spectra using SMS (Synthetic Magnitudes from Spectra), as part of the S3 package (Inserra et al. in prep). All spectra are then calibrated such that synthetic magnitudes match the photometric observations taken on the same night. We note that all raw data taken via the PESSTO programme on the NTT are publicly available in the ESO archive immediately. The reduced data are released on an annual basis in formal data releases via the Science Archive Facility at ESO. All PESSTO\footnote{Details available from  www.pessto.org} reduced data products are made available both through ESO and WISeREP \citep{wiserep}. 
%

\section{Analysis}
\label{sect:analysis}
\subsection{Reddening}
\label{sect:red}

A commonly used method to estimate extinction for normal SNe Ia relies on the expected colour evolution \citep["Lira relation",][]{lira--relation}. For SNe Iax, there is currently no evidence of uniform evolution \citep{foley--13}, so we cannot use this technique. We note however, that the colours of SN~2015H are relatively blue (Fig.~\ref{fig:15h--colour}). 
The presence of interstellar lines due to \ion{Na}{i}, is another sign of material in the line-of-sight. Numerous caveats notwithstanding, if \ion{Na}{i} features are observed in absorption, one can nevertheless correlate the equivalent width of these  with the extinction \citep{na1d-red-3,na1d-red-1,na1d-red-2}. 
In the case of SN~2015H, the spectra show no sign of the \ion{Na}{i} D features.
Taking both of the above points together, we deem the extinction to be low, and in what follows, we correct the data for Milky Way extinction in the direction of NGC~3464 only. For this, we adopt the value provided by NED, of $E(B-V)$ = 0.048 \citep{schlafly} with $R_{V}$ = 3.1 \citep{fitzpatrick}. 
%

\subsection{Photometry}
\label{sect:photometric}

Fig.~\ref{fig:15h--lc} shows the full light curve for SN~2015H. Our light curve coverage extends from approximately two weeks before $r$--band maximum (a time we estimate to be shortly after explosion) to almost 150 days post-maximum. As described in detail below, we are thus in the fortuitous position to be able to place constraints on a full set of light curve parameters: explosion epoch, peak magnitude, epoch of maximum light, the rise time to maximum, and decline rate, which has thus far been unusual for SNe Iax.

\begin{figure}[!t]
\centering
\includegraphics[width=\columnwidth]{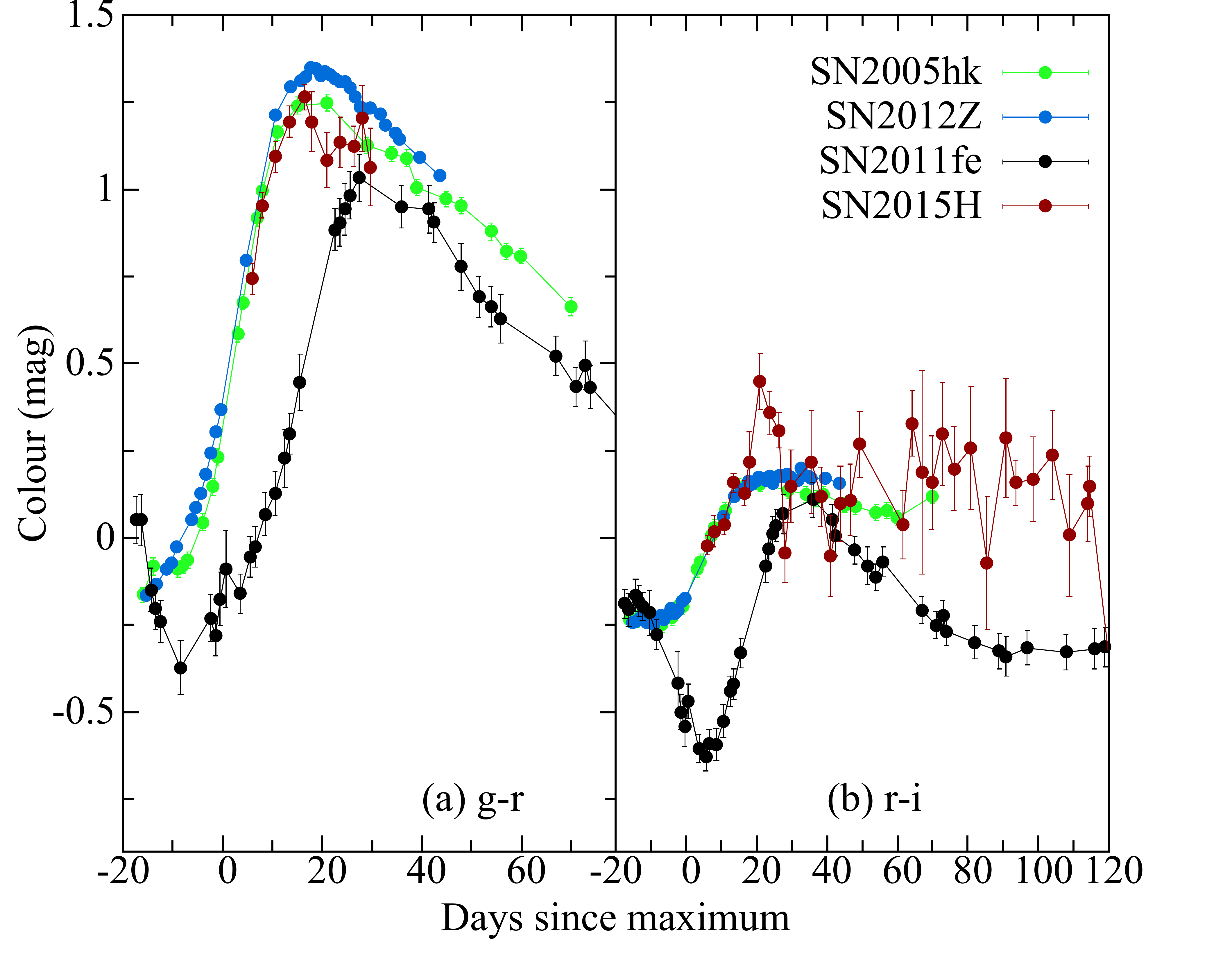}
\caption{Colour evolution of SN~2015H compared to other SNe Iax and the normal SN Ia, SN~2011fe \citep{2011fe}. In panel (a) we show the $g-r$ colour while panel (b) shows the $r-i$ colour. SN~2005hk is shown assuming negligible host extinction \citep{foley--13}. SN~2012Z and SN~2011fe have been corrected for host extinction, with data taken from \cite{comp--obs--12z} and \cite{2011fe}, respectively. Note that the colour of SN~2011fe is derived from $BVRI$ magnitudes \citep{jester--colours}. Maximum refers to the date of $r$--band maximum. 
}
\label{fig:15h--colour}
\end{figure}

\begin{figure}[!t]
\centering
\includegraphics[width=\columnwidth]{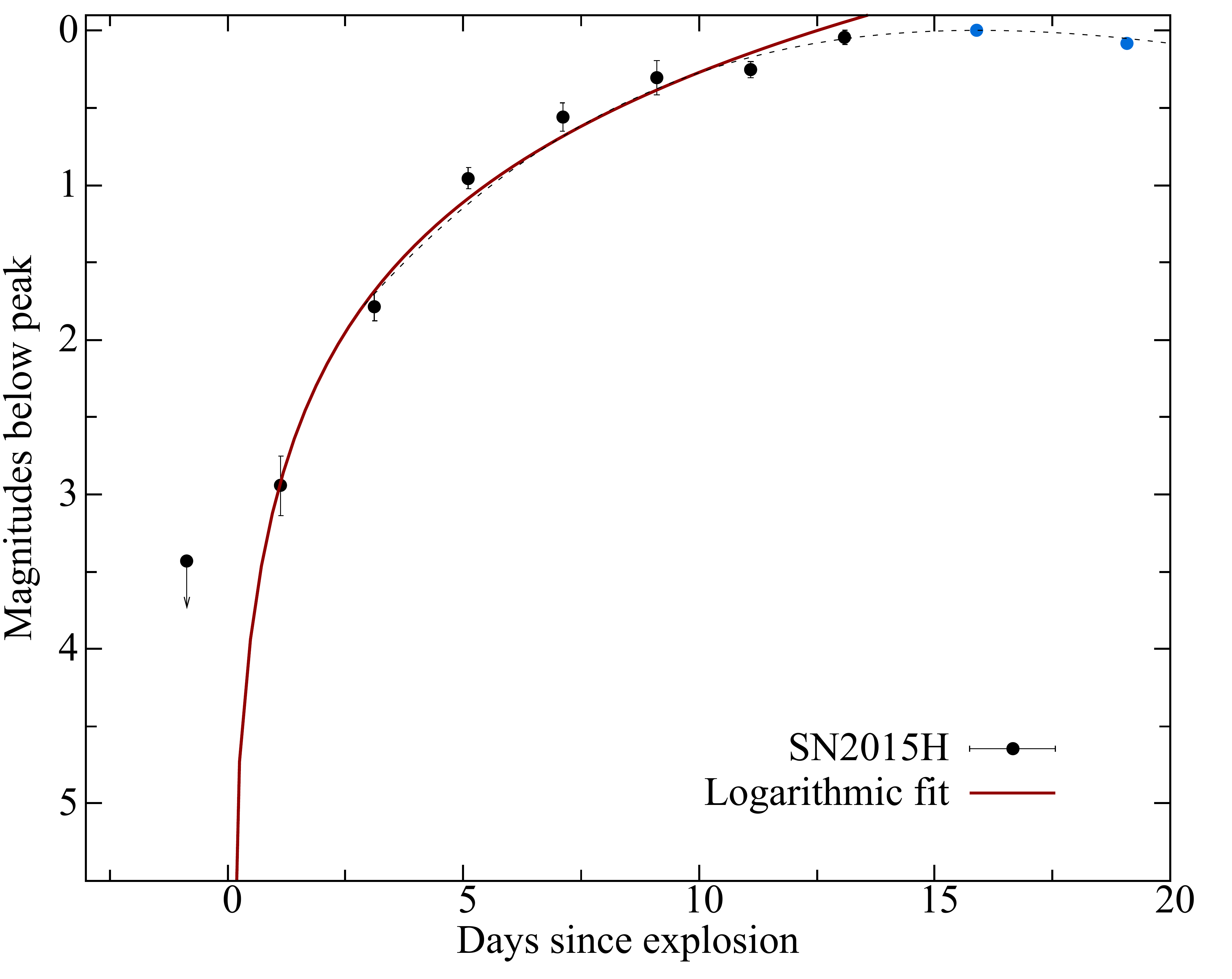}
\caption{Logarithmic fit to the rise of SN~2015H. Our pre-explosion limit is from approximately 48 hours before our first detection.
Blue points indicate points that are not included in the logarithmic fit. Our polynomial fit is shown as a dashed line.}
\label{fig:15h--exp}
\end{figure}

Beginning with our earliest images, we do not detect any source at the site of SN~2015H in images taken on or before MJD = 57045.14 (2015 Jan. 23) i.e., two days before our reported earliest detection, and over two weeks before discovery  (2015 Feb. 10). In order to place a limit on the brightness of SN~2015H at this epoch, we recovered the flux from progressively fainter artificial stars placed in the surrounding region. We find that SN~2015H would have had a  magnitude of m$_{r}$ $\ga$ 20.2 at this epoch. Studies of SNe Ia have shown that at early times the luminosity can be described by a simple power law, relating to time since explosion \cite[$L \propto$ t$^{n}$; e.g.][]{ia--rise, firth--sneia--rise}. In order to estimate the date of explosion, we fit the pre-maximum magnitudes of SN~2015H using the simple logarithmic function $m = a \log(t) + b$, and find a likely date of MJD = 57046$\pm$0.5 (Fig.~\ref{fig:15h--exp}). This date is consistent with our non-detection approximately one day earlier, despite the relatively shallow limit at that epoch. Excluding later points has no effect on our derived explosion date. Following the same method as \cite{firth--sneia--rise}, we find a rise index $n$ of $\sim$1.3. This is comparable to the broad range of values reported by \cite{firth--sneia--rise} ($\sim$1.5 to $\sim$3.5). Only a handful of SNe Iax have constraints on their explosion epoch \cite[e.g. ][for SNe 2005hk, 2008ha, 2009ku, and 2012Z, respectively]{05hk--deflag?, obs--08ha, obs--09ku, 12z--oister}, with SN~2015H being perhaps the most well-constrained to date.  
\par
To estimate the date of maximum and peak magnitude of SN~2015H, we fit the LSQ photometry with a low-order polynomial (Fig.~\ref{fig:15h--lc}). From this, we estimate that SN~2015H reached a peak absolute magnitude of $M_{r}$ = $-$17.27$\pm$0.07 on MJD = 57061.9$\pm$0.4; placing it at an intermediate brightness among the range displayed by SNe Iax, and approximately two magnitudes fainter than normal SNe Ia. 
Phases of SNe Iax are usually given relative to the date of either $B$-- or $V$--band maximum. As we lack observational data in these bands around maximum, throughout this study, all references to the epoch of maximum of SN~2015H refer to the above $r$--band date. Our light curves for SN~2015H show no signs of a secondary maximum in the SDSS $r$-- and $i$--bands, consistent with what has been previously reported for other SNe Iax \citep[e.g.][]{02cx--orig, 05hk--deflag?}. 

\par
The colour evolution of SN~2015H (in $g-r$ and $r-i$) is very similar to both SN~2005hk and SN~2012Z (Fig.~\ref{fig:15h--colour}). SN~2015H shows an increase in $g-r$ and $r-i$ colour much sooner than normal SNe Ia, similar to both SN~2005hk and SN~2012Z. In addition, SN~2015H is redder than normal SNe Ia at all observed epochs. Beginning approximately 6\,d post-maximum (the start of our $g$--band coverage), SN~2015H shows a $>$0.5 magnitude increase in $g-r$ colour over a period of approximately 10 days. After this point, the $g-r$ colour shows a steady decrease but remains redder than normal SNe Ia by at least $\sim$0.2 magnitudes at all epochs. Similarly, the $r-i$ colour of SN~2015H closely follows that of other SNe Iax; it increases slowly until a few weeks post maximum, at which point the colour remains approximately constant with $r-i$ $\simeq$0.2. The $r-i$ colours of SNe Iax do not show the  decrease and subsequent increase observed in normal SNe Ia at early times, and are consistently redder by  $\sim$0.2 magnitudes after approximately 40 days post--$r$--band maximum. 
\par
Motivated by our well-constrained explosion epoch and epoch of maximum light, we now investigate the rise time distribution of SNe Iax. For normal SNe Ia, the rise times are determined by the rate at which energy is deposited by radioactive $^{56}$Ni and diffusion through the ejecta. Most SNe Iax, however, are not well observed at these early epochs and therefore rise times exist for only a few events. We determine a rise time to $r$--band maximum for SN~2015H of 15.9$\pm$0.6\,d.
As described below, for other SNe Iax, we either take the value of the rise time quoted in the literature, or estimate a value based on available data.
For SN~2005hk, \cite{05hk--deflag?} estimate that the explosion occurred approximately 15$\pm$1\,d before $B$--band maximum, which was approximately one week before $r$--band maximum. We therefore estimate that SN~2005hk took approximately three weeks (21.8$\pm$1.2 days) to reach $r$--band maximum. \cite{obs--09ku} find that SN~2009ku had a rise time of 18.2$\pm$3.0\,d\footnote{Based on a $gri$ correction derived by applying a stretch factor to the light curve of SN~2005hk.} 

\begin{figure}[!t]
\centering
 \includegraphics[width=\columnwidth]{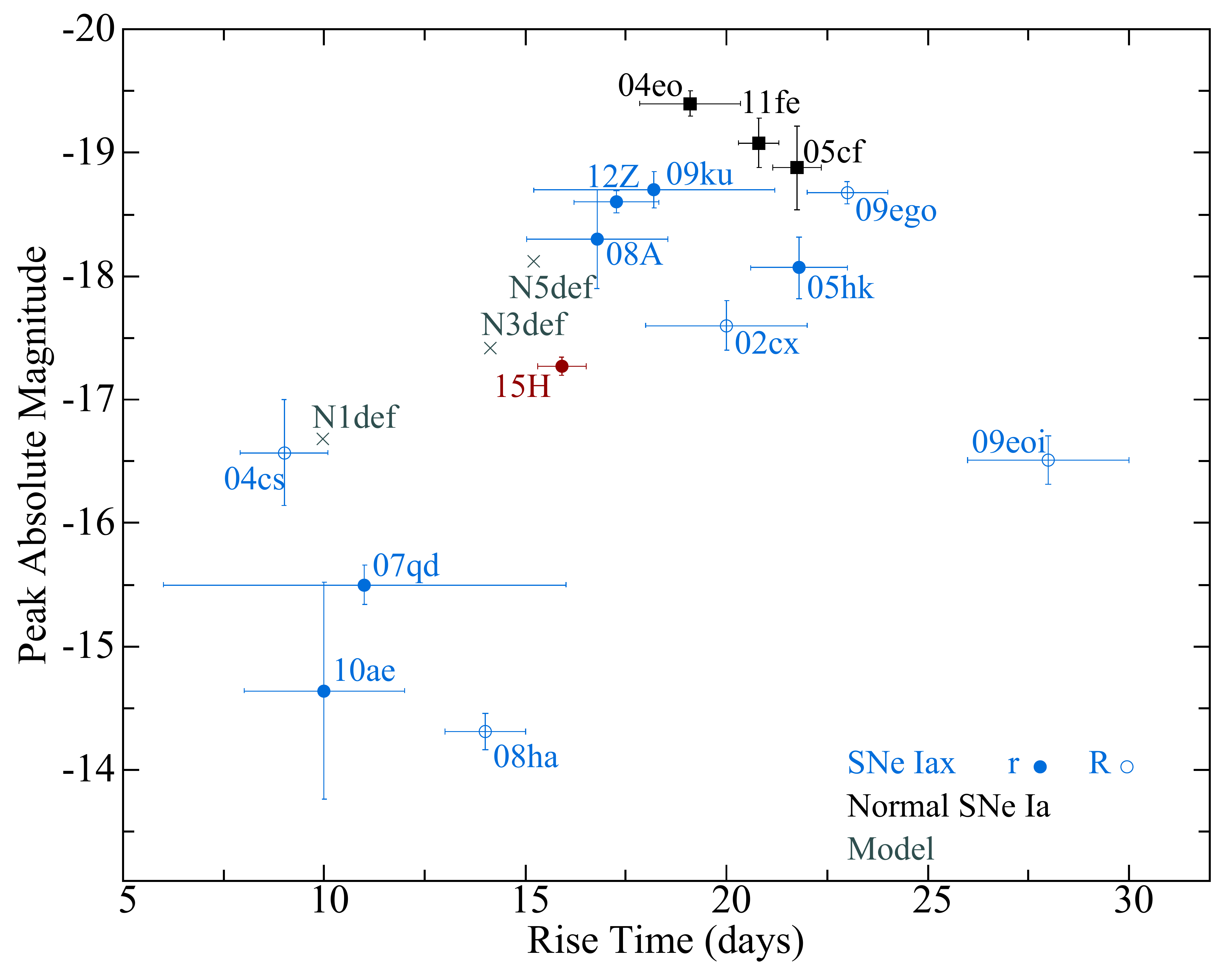}
\caption{Possible linear correlation between the rise time to peak, and peak absolute magnitude for SNe Iax in the $r$--band. For comparison, we also show the region occupied by a few normal SNe Ia,
using the estimated explosion epochs and data from sources listed in table~\ref{tab:objs}. There is no clear demarcation between the SNe Ia and the SNe Iax at the bright end. Note that values shown for SN~2005cf and SN~2011fe are based on filter transformations \citep{jester--colours}. Unfilled circles denote measurements taken in the $R$--band, with the exception of SN~2004cs, which is based on unfiltered imaging. The data sources for all objects are listed in table~\ref{tab:objs}. 
Grey crosses denote the models discussed in \S\,\ref{sect:finkcomp}. 
}

\label{fig:15h--rise}
\end{figure}
\begin{figure}[!t]
\includegraphics[width=\columnwidth]{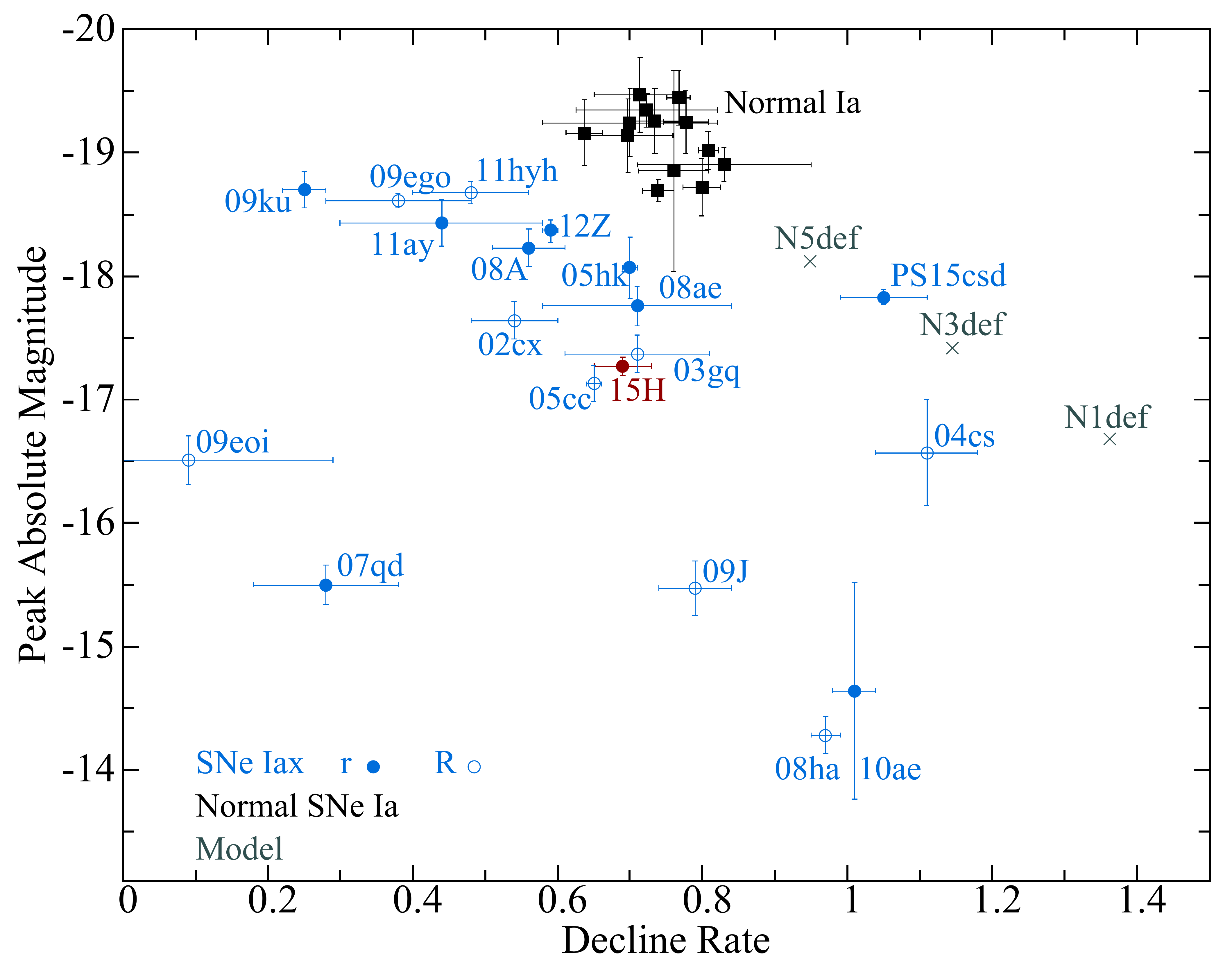}
\caption{Decline rate ($\Delta m_{15}$) versus peak absolute $r$--band magnitude for SNe Iax. As in Fig. \ref{fig:15h--rise}, unfilled circles are SNe Iax in the $R$--band, again with the exception of SN~2004cs, which is based on unfiltered imaging. Grey crosses denote the models discussed in \S~\ref{sect:finkcomp}. Black points are values typical of normal SNe Ia, taken from Carnegie Supernova Project\protect\footnotemark ~\protect\citep[][]{csp--survey} fits to SNe Ia data. 
Note objects shown here that are not shown in Fig.~\ref{fig:15h--rise} as they lack good pre-maximum coverage are: SNe 2003gq, 2005cc, 2008ae, 2011ay, PTF~11hyh, and PS15csd.  Distances to PS15csd, and PTF 11hyh, 09ego, and 09eoi are derived based on the estimated redshift. The data sources for all objects are listed in table~\ref{tab:objs}.
}
\label{fig:15h--decline}
\end{figure}
\footnotetext{http://csp.obs.carnegiescience.edu}

For SNe 2008A and 2012Z, we use the pre-maximum $r$--band photometry presented in \cite{08a--lc} and \cite{comp--obs--12z}, respectively, and find rise times of 16.8$\pm$1.8\,d for SN~2008A, and 17.3$\pm$1.1\,d for SN~2012Z, respectively. SN~ 2010ae was undetected up to five days before discovery \citep{obs--10ae}.
Using this limit, combined with data from \citet{obs--10ae}, we estimate a rise time to $r$--band maximum of 10$\pm$2\,d. Based on limited pre-maximum detections of SN~2007qd \citep{obs--07qd}, we estimate a rise time to $r$--band maximum of 11$\pm$5 d.
In addition to the above objects, we examined the unfiltered light curve of SN~2004cs presented by \cite{foley--13} and found a rise time of $\sim$9.0$\pm$1.1 days. 
In order to estimate the rise time of SN~2008ha, \cite{obs--08ha} stretched the light curve of 
SN~2005hk, and found a rise time of $\sim$10 days to $B$--band maximum. $R$--band maximum occurred a further four days later, and we therefore assume a rise time for SN~2008ha of 14$\pm$1\,d.
Finally, we estimate rise times to $R$--band maximum for PTF~09ego (23$\pm$1 days) and PTF~09eoi (28$\pm$2 days), based on data from \citet{slow--ptf}. As shown in Fig.~\ref{fig:15h--rise}, those SNe Iax that are brighter, tend to exhibit longer rise times to peak than those that are fainter. The rise times and peak luminosities of SNe Iax in the $r$--band show a range of values up to what is typical for normal SNe Ia, with PTF~09eoi being a clear outlier.

\begin{figure*}
\centering
\includegraphics[width=\textwidth]{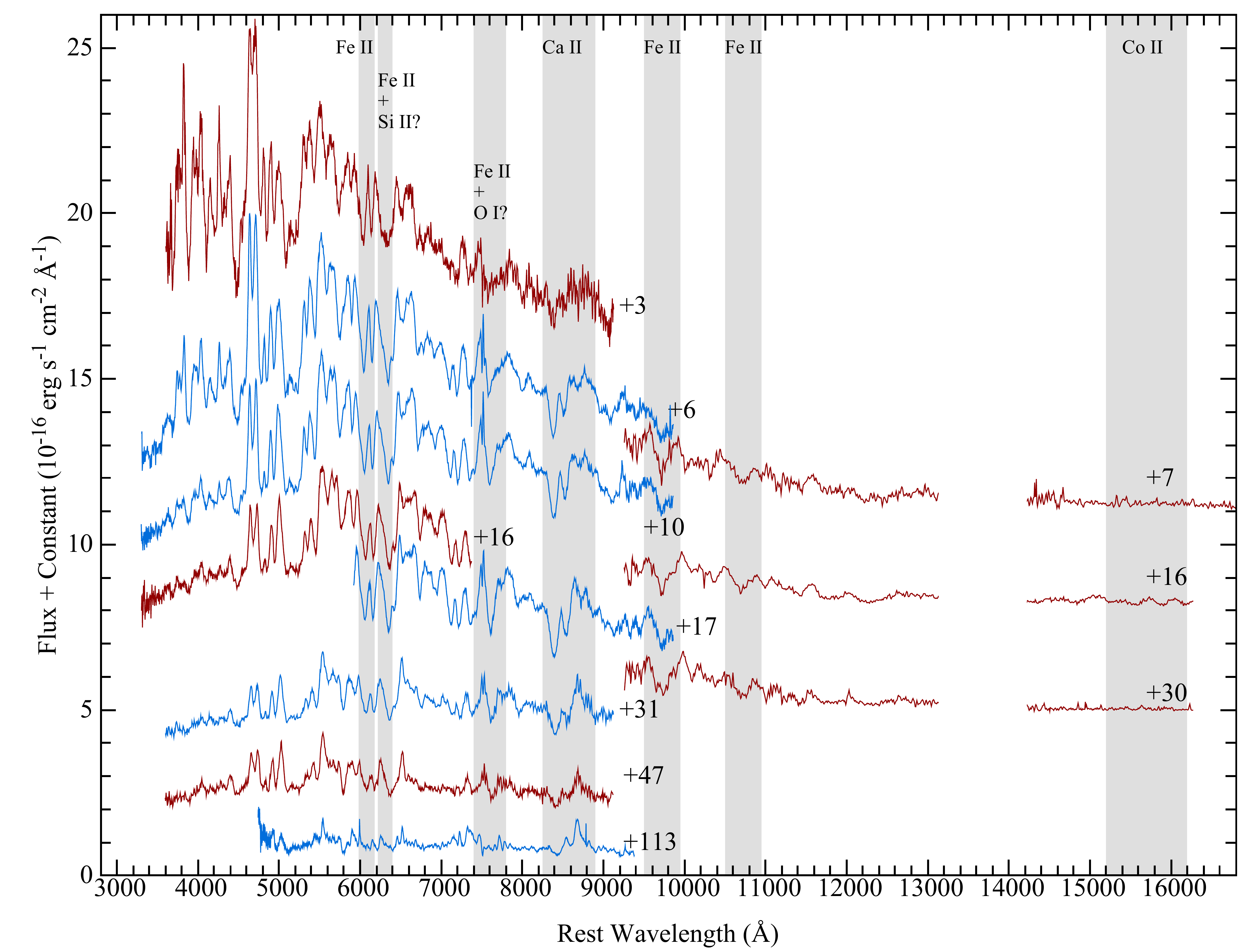}
\caption{
Spectroscopic sequence of SN~2015H. Spectra have been corrected for redshift and galactic extinction. Epochs are given relative to maximum light at MJD = 57061.9. Regions showing strong telluric features in the NIR have been omitted. Near-IR spectra have also been smoothed, to highlight the features present. We also highlight the features discussed in Section~\ref{sect:spect-analysis} and those present throughout the spectroscopic evolution which were used for velocity measurements in Fig.~\ref{fig:15h--velocities}. Arbitrary vertical offsets, and different colours for alternating spectral phases have been used for clarity.}
\label{fig:15h--spec}
\end{figure*}

\par
We now turn our attention to the post-maximum evolution of SN~2015H. In Fig.~\ref{fig:15h--decline}, we show the relation between peak magnitude and decline rate observed in SNe Iax and normal SNe Ia which has also been considered in previous studies \citep[e.g.][]{obs--09ku, foley--13, obs--10ae}. We use the standard definition of decline rate: i.e., the change in observed magnitude in a given filter between maximum light and 15\,d thereafter. For SN~2015H, we find $(\Delta m_{15})_{r}$ = 0.69$\pm$0.04 which is similar to that observed in some normal SNe Ia, despite being approximately two magnitudes fainter. It is also consistent with the roughly linear relation observed in SNe Iax (Fig.~\ref{fig:15h--decline}), although PTF~09eoi is again a clear outlier, as is SN~2007qd in this instance.
Together, Figs.~\ref{fig:15h--rise} and \ref{fig:15h--decline} show that SNe Iax with longer rise times show slower decline rates, indicating that the width of a SN Iax light curve is indeed linked with its absolute magnitude and tied to $^{56}$Ni production.
\par
We test whether the trends observed in Figs.~\ref{fig:15h--rise} and \ref{fig:15h--decline} represent statistically significant correlations via the use of the Pearson correlation coefficient, which  is a measure of whether a linear correlation exists between two variables. A value of $\pm$1 represents a complete correlation, while 0 indicates no correlation.
For the decline rate versus peak absolute magnitude, we find a correlation coefficient of 0.39 (p-value $\sim$0.09) when considering the entire sample. Excluding SN~2007qd and PTF~09eoi\footnote{The nature of PTF~09eoi may be uncertain: its limited spectral series shows some similarities to SNe Iax at early epochs, but the matches worsen with time \citep{slow--ptf}. SN~2007qd does not have spectra beyond $\sim$15 d post $B$--band maximum \citep{obs--07qd}. Both PTF~09eoi and SN~2007qd lie in a region of M$_r$ vs. ($\Delta$m$_{15}$)$_r$ separate from the rest of the sample. Given the lack of data for these objects, we are not able to make definitive statements about their nature.} we find this value increases to 0.72 (p-value $\sim$0.001). For the rise time versus peak absolute magnitude distribution of SNe Iax, we find a correlation coefficient of $-0.52$ (p-value $\sim$0.08) when including all objects, and $-$0.71 (p-value $\sim$0.02) if PTF~09eoi and SN~2007qd are excluded. Thus, our findings are suggestive of the existence of correlations between the absolute ($r$--band) magnitude of SNe Iax with both rise time and decline rate.
\par
Using the parameters derived from our light curve, we now seek to constrain 
the amount of $^{56}$Ni produced during the explosion, and the total amount of material ejected.
Our maximum light coverage of SN~2015H unfortunately only includes one filter, so we are unable to construct a bolometric light curve. We note however, that the colour and decline rates of SN~2015H and SN~2005hk, are very similar (see Figs. \ref{fig:15h--colour} and \ref{fig:15h--decline}), and SN~2005hk has extensive pre-maximum coverage in numerous filters. We therefore stretched the light curves of SN~2005hk from \cite{comp--obs--12z} and scaled them to match SN~2015H, allowing us to use these values to construct a pseudo-bolometric light curve for SN~2015H across $ugriJH$ filters. Applying Arnett's law \citep{arnett--law}, and the descriptions provided by \cite{limit--hubble} and \cite{2002es}, we estimated the $^{56}$Ni and ejecta masses.
By taking a peak bolometric luminosity of $\log (L/L_{\odot}) = 8.6 $, and a rise time to bolometric maximum of $\sim$14.5 days, we find that SN~2015H produced $\sim$0.06 M$_{\odot}$ of $^{56}$Ni. 
Using an average ejecta velocity of $\sim$5500 km~s$^{-1}$ (\S\ref{sect:spect-analysis}, Fig.~\ref{fig:15h--velocities}) we estimate an ejecta mass of $\sim$0.50 M$_{\odot}$. These values are based on simple scaling laws appropriate for normal SNe Ia. As a test of the validity of our estimates, we used the light curve model described in \cite{cosimo--lc--code} to fit our pseudo-bolometric light curve; we find that it is best matched by $\sim$0.06 M$_{\odot}$ of $^{56}$Ni and $\sim$0.54 M$_{\odot}$ of ejecta. Thus, for SN~2015H, we find our values for the $^{56}$Ni mass and the ejecta mass to be consistent both with other estimates for SNe Iax \citep[see Fig.~15 of][]{obs--05hk--08a} and the fact that it is of intermediate brightness among this class.

\begin{figure*}
\centering
\includegraphics[width=\textwidth]{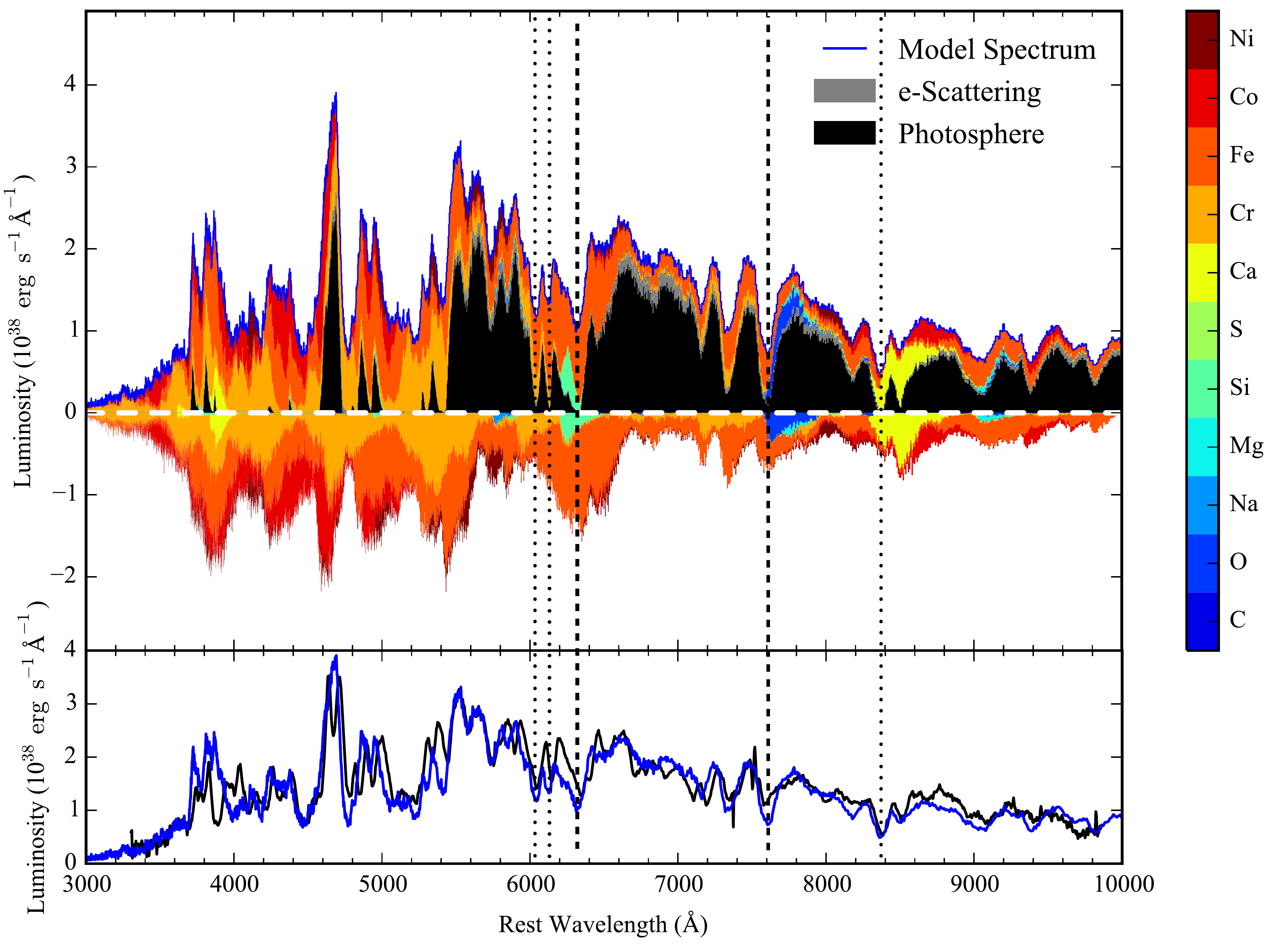}
\caption{In the bottom panel of this plot, we show a comparison of our model TARDIS spectrum, in blue, to a spectrum of SN~2015H, in black, at 6 days post $r$--band maximum. Our TARDIS model provides good agreement with many of the features observed in SN~2015H. In the top panel, we colour code a histogram of luminosity per wavelength bin to the atomic number of the element responsible for the last interaction with an escaped Monte Carlo packet. Features below/above the dashed line show the relative contribution of these elements in removing from or adding to the observed flux during the simulation (based on the analysis of interaction histories of escaping MC packets).  
Packets that do not experience line interactions within the SN ejecta are shown in either black or grey. Black indicates where packets have passed through the entire model region with no interaction, while grey packets have only experienced scattering via free electrons. Dotted vertical lines indicate features for which we measure a velocity, (shown in Fig.~\ref{fig:15h--velocities}). Dashed vertical lines show features that are discussed further in the main text. }
\label{fig:15h--tardis-id}
\end{figure*}

%

\subsection{Spectroscopy}
\label{sect:spect-analysis}

In Fig.~\ref{fig:15h--spec} we show the full series of optical and IR spectra for SN~2015H. Our first spectrum is from approximately three days post $r$--band maximum, while our last spectrum is 110 days later. The spectra of SN~2015H show features typical of SNe Iax i.e., dominated by IGEs with line velocities lower than those typically measured in normal SNe Ia by a factor of $\sim$2.
\par
In order to facilitate line identifications, we compare our spectra to calculations made with the 1D Monte Carlo radiative transfer code TARDIS\footnote{http://tardis.readthedocs.org/en/latest/} \citep{tardis}. For a given set of input model parameters, TARDIS computes a self-consistent description of the SN ejecta and emergent spectrum.
TARDIS is described in detail by \cite{tardis}, while details of the model adopted here are discussed in Appendix~\ref{apdx:tardis}. 
\par
We note that the photospheric approximation employed by TARDIS (i.e. a sharp boundary between optically thick and thin regions) is often used to describe optical wavelengths and leads to synthetic spectra in reasonable agreement to results of other, more sophisticated, radiative transfer codes \citep[][]{tardis}. This approximation however, is limited when describing the IR regime where opacities are substantially lower.
Accordingly, we limit our TARDIS comparison to optical wavelengths. We carry out our TARDIS modelling on our +6\,d spectrum only: between +3\,d to $\sim$+30\,d, the spectral features show little evolution in shape and strength, possibly indicating that the ejecta is well mixed; furthermore, our +6\,d spectrum has a higher signal-to-noise compared to the earlier (+3\,d, classification) spectrum. Beyond epochs of $\sim$30\,d, the assumptions inherent in the TARDIS model (e.g. definition of the photosphere)
are no longer robust, so we do not attempt to model spectra taken at these epochs.
\par
In Fig.~\ref{fig:15h--tardis-id} we show a synthetic spectrum that indicates the elements responsible for producing the main spectral features. In order to interpret the spectrum, we show a histogram colour-coded by the atomic number of the element responsible for the last interaction experienced by Monte Carlo packets that escaped from the model region (i.e., those packets that are responsible for contributing to the emergent spectrum). Fig.~\ref{fig:15h--tardis-id} shows that the synthetic spectrum is able to reproduce many of the features observed in SN~2015H. From the colour-coded histogram, it is clear that the spectrum is dominated by interactions with IGEs. Below we discuss a few specific cases of elements that have been reported in other SNe Iax. 

\par
\ion{Si}{ii} $\lambda$6355, one of the characteristic features of normal SNe Ia, has been reported in a number of SNe Iax \cite[e.g. ][]{05hk--deflag?, obs--07qd, obs--10ae} at epochs extending from pre-maximum to approximately two weeks post $V$--band maximum. Other features attributed to \ion{Si}{} have also been reported in some SNe Iax at early times \cite[e.g. \ion{Si}{iii} $\lambda$4553 and \ion{Si}{ii} $\lambda$5972; ][]{read--02cx--spectra, early--late--08ha--obs, 12z--oister}.
About two weeks post-maximum however, the \ion{Si}{ii} features blend with features due to \ion{Fe}{ii} and cannot be easily distinguished \citep{read--02cx--spectra}.
As our spectroscopic campaign only began post--$r$--band maximum we cannot identify clear features due to \ion{Si}{ii} in our spectra. This is borne out in Fig.~\ref{fig:15h--tardis-id}, where the complex broad feature around $\sim$6300\,\AA ~is a likely blend of \ion{Fe}{ii} and \ion{Si}{ii}. Features near 4550 and 5970\,{\AA} can be attributed solely to IGEs for the spectrum under consideration.
\par
Unburned oxygen in the ejecta has been predicted by models of deflagrations of WDs  \citep{ia--antideflag,kicked--remnants,3d--deflag--sim--obs}. Features reported as \ion{O}{i} $\lambda$7773 have been identified in some SNe Iax \citep{02cx--late--spec, 05hk--400days, obs--08ha, obs--09ku}. It has also been claimed however, that similar features in the early spectra of SN~2005hk and late time spectra of SN~2002cx and SN~2008A can be explained by \ion{Fe}{ii} \citep{05hk--deflag?, obs--05hk--08a}. Our spectra show a broad absorption feature around $\sim$7600 \AA, which our model indicates can be explained by a strong blend of \ion{Fe}{ii} and \ion{O}{i}. Similar to the case of \ion{Si}{ii} $\lambda$6355 above, we do not identify this feature as being solely due to \ion{O}{i}, but instead favour an interpretation where \ion{Fe}{ii} is the main contributor. We reiterate that features such as \ion{Si}{ii} and \ion{O}{i} may well be easier to identify at earlier epochs.

\begin{figure}
\centering
\includegraphics[width=\columnwidth]{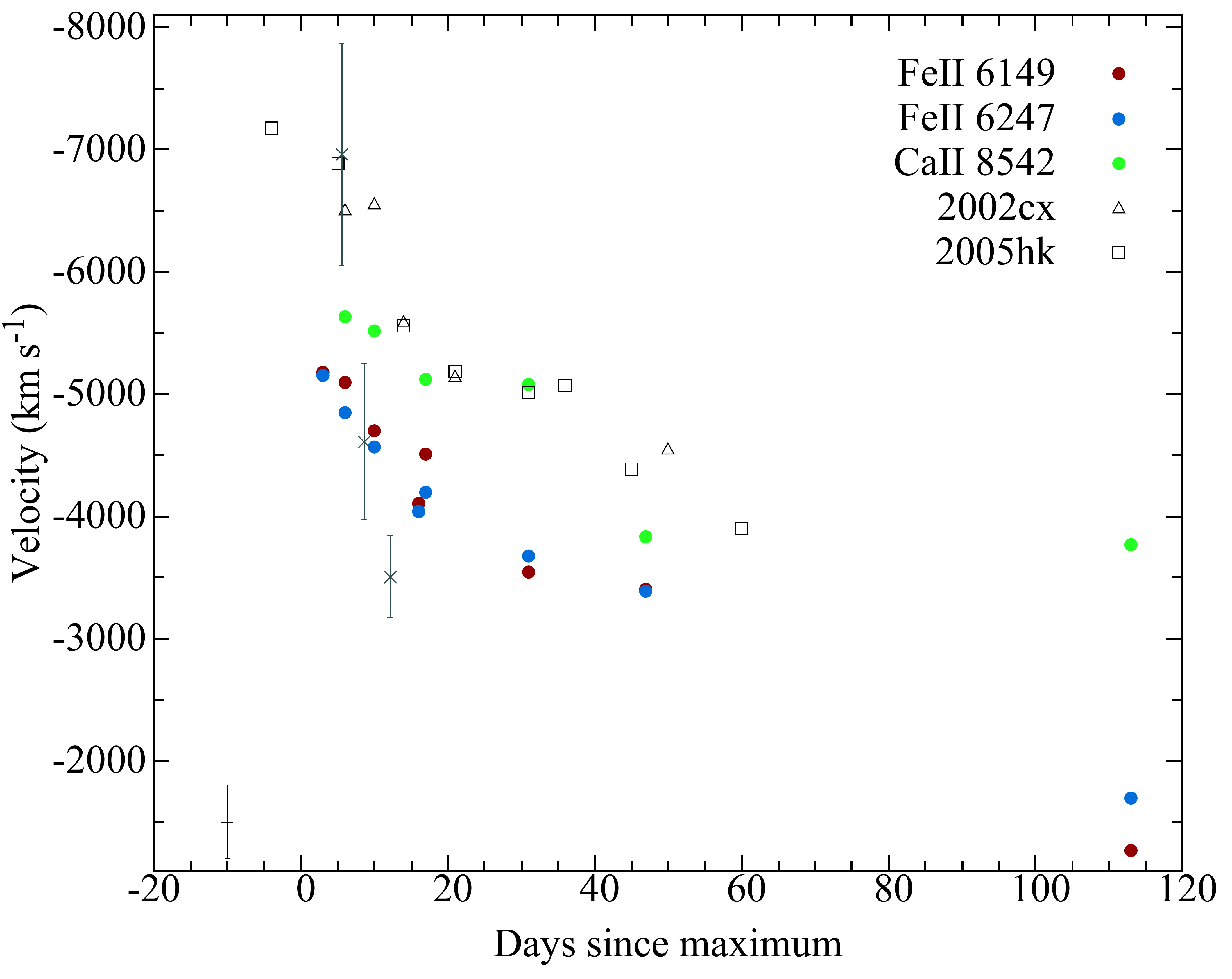}
\caption{Evolution of absorption minima for SN~2015H for a selection of lines that show minimal blending. For comparison, we show velocities measured for \ion{Fe}{ii} $\lambda$6149 for both SN~2002cx and SN~2005hk. Given in black is a typical error bar for measured velocities. Points marked with grey crosses indicate velocities of \ion{Fe}{ii}\,$\lambda$6247 measured for the N3def model (see \S~\ref{sect:finkcomp}), at the three epochs shown in Fig.~\ref{fig:15h--finkcomp}. Measurements of velocity will be affected, to some degree, by Monte Carlo noise present in the model. We, therefore, estimate a median velocity of these three epochs $\pm$2\,d, with the uncertainty given by the standard deviation of these measurements. 
}
\label{fig:15h--velocities}
\end{figure}

Having identified the dominant ions shaping our optical spectrum, we now wish to identify the dominant species 
responsible for the near-IR flux. While the TARDIS model discussed above is unsuitable for this
purpose, the more sophisticated simulations discussed further in \S\ref{sect:finkcomp} can be used for this purpose. Based on comparing our spectra to these models, we find that as for the optical spectrum, our $1-1.6$\,$\mu$m spectrum taken at +16\,d is dominated by IGEs, in particular \ion{Fe}{ii} features and weak \ion{Co}{ii}  (Fig. \ref{fig:15h--spec}).
\par
As with SNe Ia, the detection of H or He in spectra of SNe Iax could be key to shedding light on the progenitor
channel(s). Indeed, the recent intriguing detection of the progenitor of the SN Iax 2012Z, has characteristics reminiscent of that of some He novae \citep{12z--prog}. 
Possible detections of \ion{He}{i} ($\lambda$5876, $\lambda$6678, $\lambda$7065) have in fact been reported for two SNe Iax: SNe~2004cs\footnote{SN~2004cs was originally classified as a SN II \citep{04cs--first}.} \citep{foley--13} and 2007J\footnote{SN~2007J was originally classified as being similar to SN~2002cx \citep{07j--cbet--1}. The subsequent development of \ion{He}{i} features caused this classification to be revised to a SN Ib \citep{07j--cbet--2}. \cite{obs--08ha} however, consider SN~2007J to be a member of the SNe Iax class. } \citep{obs--08ha}, although whether these two objects are {\it bona fide\/} members of this class is contentious \citep{slow--ptf}.

The atomic energy levels associated with the optical transitions of \ion{He}{i} are difficult to excite, so the lack of  strong \ion{He}{i} features in optical spectra is not surprising, and does not necessarily imply that it is not present in the progenitor system \citep[e.g.][]{hidden-he}. Currently, we cannot comment further on optical features (due to the lack of a sufficiently sophisticated treatment of non-thermal excitation/ionization of He in our models), but we note that at the epoch under consideration, features due to IGEs dominate the spectrum.

Strong \ion{He}{i} transitions do exist however, at near-IR wavelengths e.g., the 1s2s~$^3$S -- 1s2p~$^3$P transition at 10830\,\AA. Many SNe Iax lack observations at these wavelengths; for SN~2015H, not only do we cover this region, but our three near-IR spectra were taken at epochs comparable to those of SN~2007J for which the putative \ion{He}{i} feature was reported to be growing stronger with time \citep{07j--cbet--2}. In SN~2015H, we find a broad feature around $\sim$10700\,\AA, but the inferred velocities are too low when compared with other species to plausibly identify this feature as \ion{He}{i}. Based on models described in \S\ref{sect:finkcomp}, we find good agreement for this feature with \ion{Fe}{ii}, and consider this to be a more plausible identification. 

\par
We estimated the ejecta velocity by fitting Gaussian profiles to the blueshifted absorption minima of features that show minimal blending. Velocities observed in SNe Iax at maximum light range from $\sim$2000 km~s$^{-1}$ to $\sim$8000 km~s$^{-1}$ \citep{foley--13}, and are consistently lower than observed in normal SNe Ia by a factor of $\sim$2. Fig.~\ref{fig:15h--velocities} shows the velocity evolution of SN~2015H as measured from a selection of atomic lines that exhibit minimal blending. These velocities show a range from $\sim$5000 km~s$^{-1}$ to $\sim$6000 km~s$^{-1}$ at approximately 3\,d post-maximum. 
\ion{Fe}{ii} $\lambda$6149 and $\lambda$6247\,{\AA}
in particular show a roughly linear decrease from $\sim$5000 km~s$^{-1}$ to $\sim$1500 km~s$^{-1}$ over the observed time frame. In Fig.~\ref{fig:15h--velocities}, we also show the velocity evolution as measured from \ion{Fe}{ii} $\lambda$6149 for SNe~2002cx and 2005hk; both display higher velocities than SN~2015H, which is consistent with the latter having a slightly fainter peak absolute brightness. For the first 50\,d post-maximum light, SN~2015H displays a velocity gradient of $\sim$40 km~s$^{-1}$~d$^{-1}$. SN~2002cx and SN~2005hk show gradients of $\sim$45 km~s$^{-1}$~d$^{-1}$ and $\sim$50 km~s$^{-1}$~d$^{-1}$, respectively (Fig.~\ref{fig:15h--velocities} and \S\ref{sect:comp}). These values are similar to those seen in low-velocity gradient SNe Ia \citep{sneia--vel}. 
\par

%

\section{Comparison with SN~2002cx and SN~2005hk}
\label{sect:comp}

\begin{figure*}
\centering
\includegraphics[width=\textwidth]{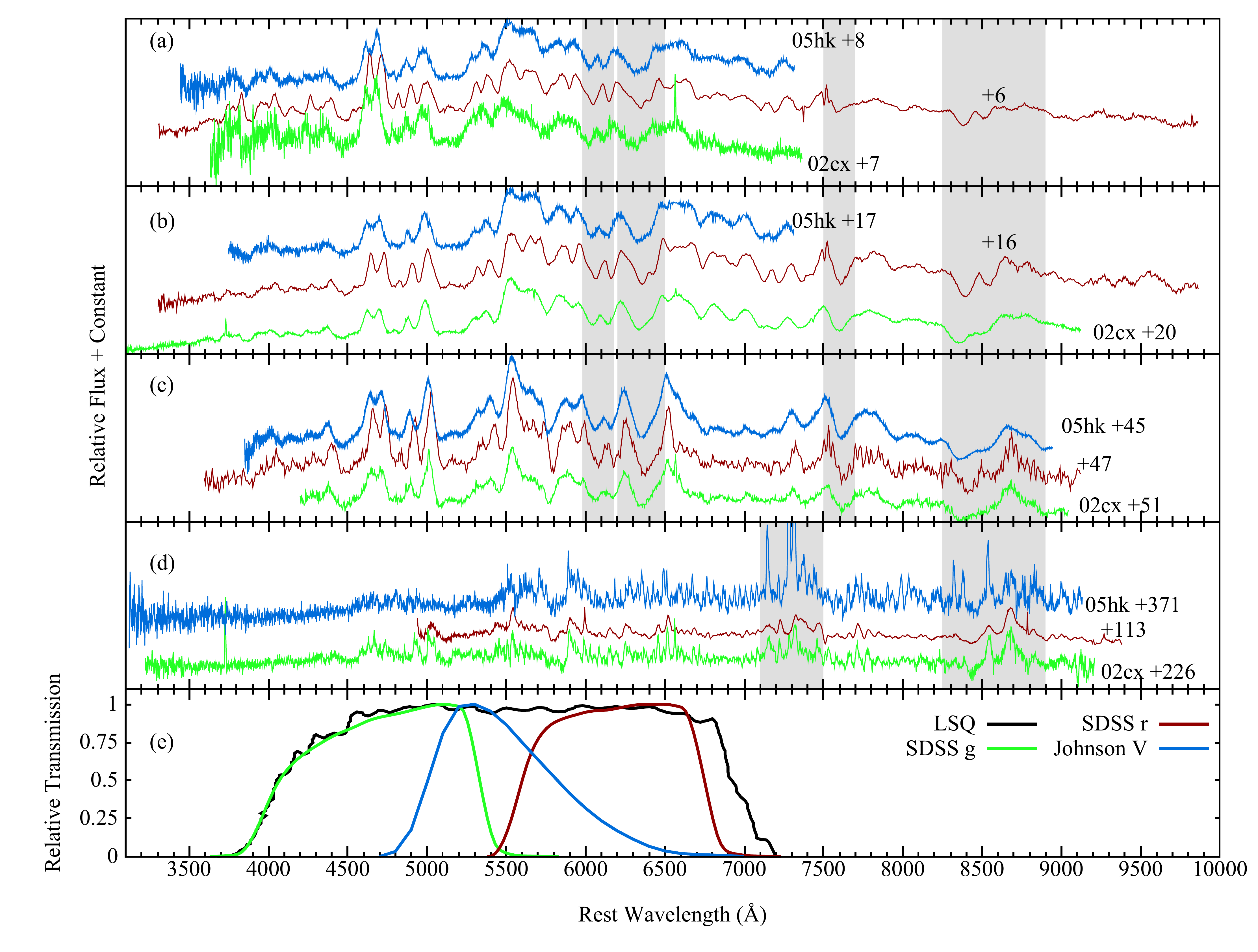}
\caption{Comparison of SN~2015H to other SNe Iax at similar epochs.
Spectra are shown in the rest frame of the host galaxy and corrected for galactic extinction. In panel (e) we show the filter function for the LSQ wide filter, along with standard filters SDSS $g$ and $r$, and $V$. Note the we scale each filter such that peak transmission has a value of 1.
Spectra of SN~2002cx and SN~2005hk were obtained from WISeREP\protect\footnotemark ~\protect\citep{wiserep} and originally from the following sources:
SNe 2002cx: \cite{02cx--orig, 02cx--late--spec}. 2005hk: \cite{05hk--blondin, 05hk--deflag?, silverman--ia}.
}
\label{fig:15h--speccomp}
\end{figure*}

In Fig.~\ref{fig:15h--speccomp} we show a selection of SN~2015 spectra spanning $\sim$\,+6 to +113\,d and compare these to two other well-observed SNe Iax (SNe 2002cx and 2005hk) at similar epochs, with comparable peak absolute brightnesses and decline rates. 
\par
Fig.~\ref{fig:15h--speccomp} shows that many of the spectral features present in both SN~2002cx and SN~2005hk are also present in SN~2015H. The most noticeable difference between these objects is their expansion velocities, with SN~2015H having a slightly lower velocity (by approximately 1000 -- 2000 km~s$^{-1}$, see Fig.~\ref{fig:15h--velocities}). Indeed, the somewhat narrower spectral
features of SN~2015H allows for the identification of features that were blended in SN~2002cx and SN~2005hk. Notably, the broad feature around 6300\,\AA ~(which we believe to be dominated by \ion{Fe}{ii}, see \S~\ref{sect:spect-analysis}) in our +6 and +16 day spectra shows clear signs of a double (and possibly triple) peak, while both SNe~2002cx and 2005hk show smoother features at these epochs. The spectra also contain permitted lines e.g., \ion{Ca}{ii} IR triplet, which could easily be identified throughout the SN evolution.
It appears as a single, broad absorption feature in this region for SN~2002cx and SN~2005hk, while in SN~2015H clear multiple absorption features belonging to this transition are evident. The lower expansion velocities measured in SN~2015H, together with the fact that it was fainter than both comparison SNe indicates that it was likely a less energetic explosion. 
\par
\footnotetext{http://wiserep.weizmann.ac.il}The similarity between the spectra of all three objects, shown in Fig.~\ref{fig:15h--speccomp}(a), (b), and (c), indicates that, despite the different explosion strengths and luminosities, they evolve at a roughly similar rate. Additionally, the similarities between our +115\,d spectrum, 
and the much later spectra of SNe~2002cx and 2005hk (+226 and +371\,d post maximum, respectively) are shown in Fig.~\ref{fig:15h--speccomp}(d)\footnote{We have chosen to compare spectra at these epochs, as no spectra of SNe~2002cx and SN~2005hk are available at comparable epochs. The spectrum presented here for SN~2002cx is the earliest spectrum available after $\sim$50\,d post-maximum, while the spectrum for SN~2005hk is the earliest spectrum available after $\sim$70\,d post-maximum.}. 
The similarities between these spectra at such different epochs for objects of a similar decline rate indicates that the spectral features of SNe Iax do not undergo much evolution throughout this period. These late phase spectra show a complex blend of forbidden lines at $\sim$7300\,\AA\, specifically due to [\ion{Fe}{ii}], [\ion{Ca}{ii}], and [\ion{Ni}{ii}] \citep{02cx--late--spec}.

%

%

\section{Comparison to Deflagration Models}
\label{sect:finkcomp}

\begin{figure}[!t]
\centering
\includegraphics[width=\columnwidth]{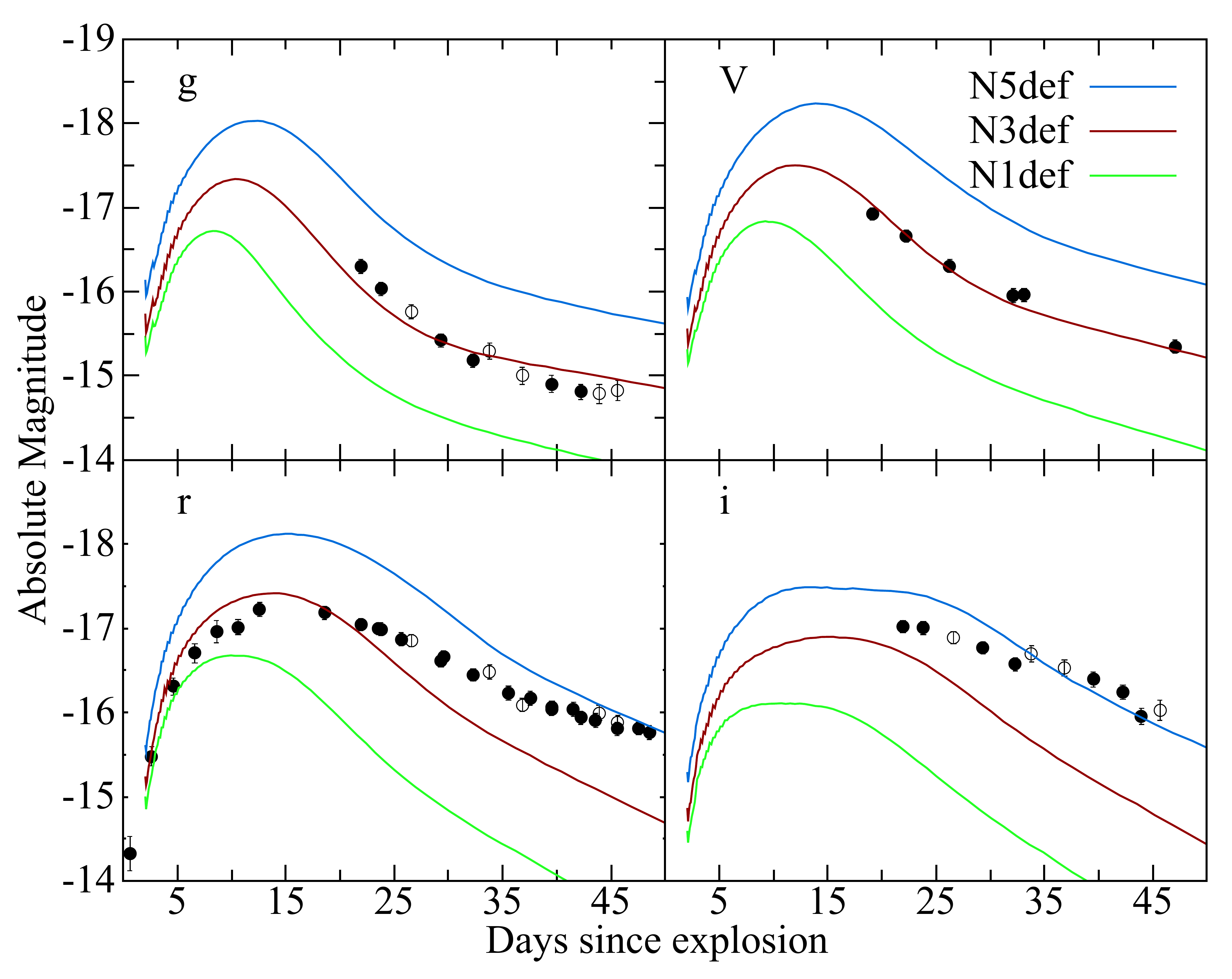}
\caption{Synthetic light curves of the N1def, N3def, and N5def models are shown in different filters. Black points are observations of SN~2015H, as in Fig.~\ref{fig:15h--lc}. The explosion date of SN~2015H is estimated to be MJD = 57046. 
}
\label{fig:15h--finklc}
\end{figure}

Having shown the similarities of SN~2015H to other SNe Iax, we now further investigate whether deflagration models proposed for this class of object are also consistent with observations of SN~2015H. In particular, we compare our observations to the models published by \cite{3d--deflag--sim--obs} and \cite{3d--deflag--sim--rem}, which have been shown to provide a reasonable match to SN~2005hk.  These models describe the pure deflagration of Chandrasekhar mass CO WDs with hydrodynamics followed from the onset of thermonuclear runaway, which is initiated by inserting multiple spherical ignition spots positioned off-centre within the WD. 
The subsequent propagating flame is treated as a sharp boundary separating the burned and unburned regions using a levelset approach \citep{level--set}. The hydrodynamic simulation of the explosion is followed until \textit{t} = 100~s by which point the ejecta are close to following homologous expansion. Ejecta density and abundances are then used in the radiative transfer calculation with the 3D Monte Carlo radiative transfer code ARTIS \citep{multid-rt-snia, artis}. Full details of the hydrodynamics, nucleosynthesis, and radiative transfer calculations are given by \cite{3d--deflag--sim--obs}.
\par
\cite{3d--deflag--sim--obs} found that the energy released during the explosion and the luminosity scale as a function of the number of ignition spots used in the model; more ignition spots lead to the release of more energy. The number of ignition spots in a simulation can therefore be used as a numerical parameter to control the strength of the explosion in the model sequence (but we note that the number of ignition points used should not necessarily be taken literally as a requirement on the ignition configuration). 
Given that SN~2015H was at least half a magnitude fainter than SN~2005hk, we limit our model comparison to those models with only a few ignition spots, as was done for SN~2005hk by \cite{3d--deflag--sim--rem}.
\par
Each model is named corresponding to the number of ignition spots (e.g. N10def uses 10 spots) used to initiate the explosion. \cite{3d--deflag--sim--obs} used a total of 14 models, from a single ignition spot up to 1600 spots. Models with few ignition spots ($\leq$100) do not release sufficient energy during the explosion to completely unbind the WD. This results in those models leaving behind a (potentially massive) bound remnant, up to $\sim$1.3 M$_{\odot}$. In addition, those models show more complex ejecta structures due to the asymmetric flame propagation. These effects, however, are relatively minor: for those objects with only a few ignition spots, there is only a modest ($\lesssim$0.2 magnitude) scatter in the synthetic $V$--band light curves across the entire range of observer orientations \citep{3d--deflag--sim--obs}. Furthermore, this scatter is even smaller in the redder bands. Consequently, we focus our comparisons on the angle-averaged observables only.

\par
In Fig.~\ref{fig:15h--finklc}, we compare the light curves of SN~2015H to synthetically generated ones from the N1def, N3def, and N5def explosion models. These are the faintest models produced by the sequence and therefore provide the best chance of matching the lower luminosities exhibited by SNe Iax. We find that of the 3 models considered here, the best agreement is with the N3def model. In particular, there is excellent agreement between the model and observations in the $g$-- and $V$--band filters, matching both the observed peak brightness and decline rate. The agreement with the peak brightness observed in the $r$--band is also reasonable. The ability of the N3def model to reproduce the brightness observed in SN~2015H indicates that the amount of $^{56}$Ni during the explosion is similar to that inferred from the observations. In \S\ref{sect:photometric}, we discussed estimates for the amount of $^{56}$Ni and ejecta mass produced by SN~2015H; by fitting the pseudo-bolometric light curve, we found values of 0.06 and 0.54\,M$_\odot$, respectively. The N3def model is consistent with our estimate of the $^{56}$Ni mass in that it yielded $\sim$0.07 M$_{\odot}$ of $^{56}$Ni. The corresponding ejecta mass is 0.2\,$M_\odot$. We discuss reasons for this mismatch below.

\par
We compare the spectra of SN~2015H to synthetic spectra generated from the N1def, N3def, and N5def explosion models in Fig.~\ref{fig:15h--finkcomp}. As with the light curves (Fig.~\ref{fig:15h--finklc}), N3def again provides the best overall match to the brightness of SN~2015H. Indeed, this model is able to match many of the broader features observed in the spectra of SN~2015H e.g. the features around $\sim$5200\,{\AA}  and $\sim$6300\,{\AA} as well as many of the narrower features e.g. \ion{Fe}{ii} $\sim$$\lambda$6050\,{\AA}. $\sim$$\lambda$6150\,{\AA}, and $\sim$$\lambda$5500\,\AA.

Despite the relatively good agreement around maximum light, there are, however, notable discrepancies between the models and the data that become increasingly apparent at later epochs. This is particularly evident in the light curves around three weeks post-explosion, when the synthetic flux in the redder filters  (i.e., redward of $\sim$6700\,\AA), drops off too rapidly, as shown in Fig.~\ref{fig:15h--finklc}; in other words, the N3def model shows a decline rate that is too fast $((\Delta m_{15})_r$ $\sim$1.1) compared to the observations ($(\Delta m_{15})_r$ = 0.69$\pm$0.04).
This mismatch feeds into the ejecta mass estimate resulting from the N3def model resulting in a mismatch of close to a factor of three. This is not surprising as a model with higher ejected mass would have a longer diffusion timescale and is likely to more effectively trap $\gamma$-rays for longer, resulting in a broader light curve with a shallower decline rate. A full exploration of ignition conditions has not been carried out. Indeed, somewhat tweaked initial conditions such as the WD central density, could potentially yield a better match to the decline rate while yielding roughly the same amount of $^{56}$Ni; this warrants further investigation.

%
\section{Summary}
\label{sect:sum}

\begin{figure*}
\centering
\includegraphics[width=\textwidth]{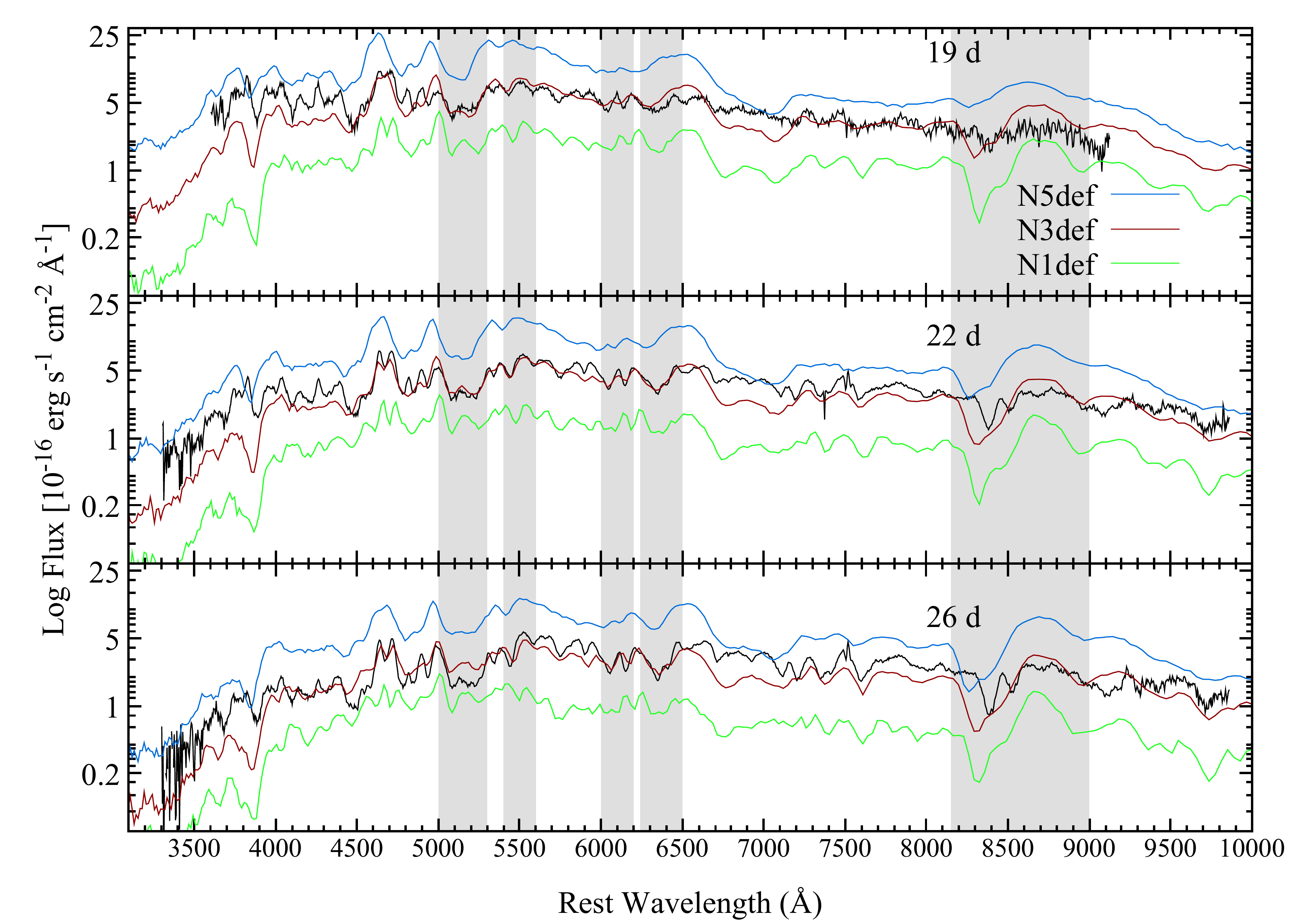}
\caption{Comparison of SN~2015H to the deflagration models of \cite{3d--deflag--sim--obs}. Observed spectra are shown in black. Comparison spectra at similar epochs are shown for the N1def, N3def, and N5def models. Dates are given relative to an explosion date of MJD = 57046.
The model fluxes are predicted an absolute scale, with no scaling factors applied.
The observed spectra have been shifted to the rest frame, corrected for reddening, and scaled to coeval photometry as described in \S\,\ref{sect:obsandreduc}.
}
\label{fig:15h--finkcomp}
\end{figure*}

We presented optical and infrared photometric and spectroscopic observations of a peculiar Type Ia SN, 2015H. Through our extensive photometric coverage of SN~2015H, we found that it peaked with an absolute  magnitude of $M_r = -17.27\pm$0.07. Pre-explosion imaging allowed us to constrain the epoch of explosion, and consequently, the rise time to maximum to approximately two weeks. The decline rate of SN~2015H is similar to that observed in some normal SNe Ia, 
in spite of being approximately two magnitudes fainter.

\par
SN~2015H shows many of the spectroscopic features typical of the SN Iax class and similarities to two of the best studied members: SNe~2002cx and 2005hk. Specifically, spectra at all epochs show low-velocity features (velocities approximately half what is observed in normal SNe Ia) and our radiative transfer modelling suggests that many of the features present in the spectra of SN~2015H can be explained by, or at least involve a significant degree of blending with, IGEs. 
\par
We compared our observations with the deflagration models of M$_{\mathrm{Ch}}$ CO WDs published by \cite{3d--deflag--sim--obs} and found good agreement, the best match being with one of their fainter models (N3def). For epochs up to approximately 50\,d post-explosion, this model accounts for the luminosity observed in the $g$-- and $V$--bands for SN~2015H.  We find good agreement between the amount of $^{56}$Ni produced by this best-matching model (0.07 M$_{\odot}$) and our estimate, based on fitting the pseudo-bolometric light curve of SN~2015H, ($\sim$0.06 M$_{\odot}$). This model is also able to reproduce many of the spectroscopic features observed in SN~2015H. 
\par
The agreement between the model and observations, however, begins to deteriorate at approximately three weeks post-explosion, particularly at wavelengths longer than $\sim$6700 \AA. While we find that N3def provides good agreement with the peak luminosity in the $r$--band, the light curve evolution of the model it too rapid; the rise time of N3def is shorter than SN~2015H while the decline rate is faster (see Figs.~\ref{fig:15h--rise} and \ref{fig:15h--decline}).
\cite{3d--deflag--sim--rem} found similar results when comparing the N5def model to SN~2005hk. The most likely cause of faster evolution is that the ejecta mass in the models is too low. Indeed, N3def produces only 0.20 M$_{\odot}$ of ejecta, while our fit to the pseudo-bolometric lightcurve of SN~2015H suggests a value close to $\sim$0.54 M$_{\odot}$. Increasing the ejecta mass would increase the diffusion time and slow the evolution, as required by the data. Within the current \citep{3d--deflag--sim--obs} sequence of models, adopting a model with higher ejecta mass would also result in increased $^{56}$Ni mass. However, the \cite{3d--deflag--sim--obs} simulations were not intended as a systematic exploration of the full parameter space for deflagration models, and there is scope for considerable further study. We note, in particular, that \cite{kicked--remnants} presented models adopting a lower WD central density and different ignition conditions from the \cite{3d--deflag--sim--obs} simulations, and obtained a lower fraction yield of IGE to total ejecta mass, suggesting that it might be possible to develop models with higher ejecta mass, without a corresponding increase in the mass of $^{56}$Ni. We note that such a model would also be likely to have lower specific kinetic energy and thus lower line velocities, which may further improve the agreement with key spectral features (see e.g. the Ca {\sc ii} near-IR triplet in Fig. \ref{fig:15h--finkcomp}.

A full study testing the parameter space of M$_{\mathrm{Ch}}$ WDs is needed to explore this further.  Similar model exploration should also be considered for alternative scenarios, such as the pulsational delayed detonation scenario: as discussed by \cite{comp--obs--12z}, this may be a good match for events as bright as
SN~2012Z, but it remains to be seen whether events as faint as SN~2008ha and SN~2010ae, or indeed even SN~2015H, can be comfortably fit within such a framework.
\par
An interesting prediction from the deflagration model we have considered is that the SN explosion will leave behind a massive ($\sim$1.2 M$_{\odot}$ for N3def) bound\footnote{As defined by \cite{3d--deflag--sim--obs}, "bound" is taken to mean gravitationally bound at the end of the corresponding hydrodynamical simulation, $t = 100$~s.} remnant. Our own estimate for the ejecta mass of SN~2015H ($\sim$0.54 M$_{\odot}$) suggests a smaller, but still substantial, remnant mass of $\sim$0.85 M$_{\odot}$ (assuming a $M_{\mathrm Ch}$ progenitor). It may be possible that such a remnant already contributes a non-negligible fraction to the observed flux  \citep{3d--deflag--sim--rem}. Indeed, this contribution may be more significant for N3def than brighter models, such as N5def, as the fraction of $^{56}$Ni contained within the bound remnant compared to the ejecta is much higher (by $\sim$40\% compared to N5def). 
Interestingly, the detection of a point-source at the location of one of the faintest SNe Iax, 2008ha has been reported over four years after the SN discovery \citep{08ha--prog}, by which time, the SN should have faded away. It is conceivable that an explosion remnant may contribute to this emission, but further investigation is needed. 
\par

Models involving the deflagration of CO WDs have previously been shown to be broadly consistent with those SNe Iax that are bright \citep[e.g. SN2005hk, ][]{3d--deflag--sim--rem}. In this study, we have shown that models of weaker deflagrations in $M_{\mathrm{Ch}}$ CO WDs producing $\sim$0.07 M$_{\odot}$ of $^{56}$Ni are able to reproduce the features observed in fainter events such as SN2015H. Combined with the hybrid CONe WD deflagration model proposed for the very faintest members of the class \citep[e.g. SN2008ha][]{cone--deflag--sim}, this suggests that deflagrations of WDs are able to account for SNe Iax across the entire brightness range of SNe Iax. Whether such deflagrations are the sole or dominant channel giving rise to SNe Iax requires further investigation, but we conclude that the models discussed here do show promise, and merit continued investigation and refinement. We also stress the need for further observations of this class of supernova, including objects that are discovered post-peak, such as SN~2015H.

%

\begin{acknowledgements}
We are grateful to C. White for providing the photometric data on PTF 09eo, 09eoi, 09eiy, and 11hyh in digital form, and U. N\"obauer for providing TARDIS analysis tools. 
RK and SAS acknowledge support from STFC via ST/L000709/1.
This work makes use of observations from the LCOGT network. Based on observations collected at the European Organisation for Astronomical Research in the Southern Hemisphere, Chile as part of PESSTO, (the Public ESO Spectroscopic Survey for Transient Objects Survey) ESO program ID 188.D-3003. 
IRS was supported by the Australian Research Council Laureate Grant FL0992131. 
The authors gratefully acknowledge the Gauss Centre for Supercomputing (GCS) for providing computing time through the John von Neumann Institute for Computing (NIC) on the GCS share of the supercomputer JUQUEEN at J\"ulich Supercomputing Centre (JSC). GCS is the alliance of the three national supercomputing centres HLRS (Universit\"at Stuttgart), JSC (Forschungszentrum J\"ulich), and LRZ (Bayerische Akademie der Wissenschaften), funded by the German Federal Ministry of Education and Research (BMBF) and the German State Ministries for Research of Baden-W\"urttemberg (MWK), Bayern (StMWFK) and Nordrhein-Westfalen (MIWF). 
Support for LG is provided by the Ministry of Economy, Development, and Tourism's Millennium Science Initiative through grant IC120009, awarded to The Millennium Institute of Astrophysics, MAS. LG acknowledges support by CONICYT through FONDECYT grant 3140566.
KM acknowledges support from the STFC through an Ernest Rutherford Fellowship.
RP acknowledges support by the European Research Council under ERC-StG grant EXAGAL-308037.
SJS acknowledges funding from the European Research Council under the European Union's Seventh Framework Programme (FP7/2007-2013)/ERC Grant agreement no. [291222]
and STFC grants ST/I001123/1 and ST/L000709/1.
WH acknowledges support by TRR 33 ‘The Dark Universe’ of the German 
Research Foundation (DFG) and the Excellence Cluster EXC153 ‘Origin and
Structure of the Universe’.
HC acknowledges support from the European Union FP7 programme through ERC grant number 320360.
This research has made use of the NASA/IPAC Extragalactic Database (NED) which is operated by the Jet Propulsion Laboratory, California Institute of Technology, under contract with the National Aeronautics and Space Administration. 
\end{acknowledgements}

\bibliographystyle{aa}

\begin{appendix}
\onecolumn
\section{Tables}
\label{apdx:tables}

\begin{table}[h!]
\centering
\caption{Local sequence stars used to calibrate SN~2015H LCOGT photometry.}\tabularnewline
\label{tab:15h--lcogt-std}\tabularnewline
\begin{tabular}{llllll}\hline
\hline
No.  & RA & DEC &  $g$ & $r$ & $i$  \tabularnewline
& &  &  (mag) & (mag) & (mag)  \tabularnewline
\hline
\hline
1 & 10:54:39.42 & -21:07:23.2 & 17.53(0.01) & 16.34(0.01) & 15.84(0.01)    \tabularnewline
2 & 10:54:59.94 & -21:08:32.1 & 16.42(0.01) & 15.68(0.01) & 15.42(0.01)    \tabularnewline
3 & 10:54:48.35 & -21:04:59.1 & 18.99(0.01) & 17.56(0.01) & 16.50(0.01)    \tabularnewline
4 & 10:55:07.80 & -21:00:34.5 & 16.75(0.01) & 16.00(0.01) & 15.76(0.01)    \tabularnewline
5 & 10:54:58.77 & -21:00:40.5 & 15.62(0.01) & 15.16(0.01) & 15.02(0.01)    \tabularnewline
6 & 10:54:45.04 & -21:01:21.9 & 15.76(0.01) & 15.25(0.01) & 15.10(0.01)    \tabularnewline
7 & 10:54:19.14 & -21:02:44.2 & 17.15(0.01) & 15.86(0.01) & 15.32(0.01)    \tabularnewline
8 & 10:54:31.45 & -21:03:53.3 & 17.33(0.01) & 16.35(0.01) & 15.99(0.01)    \tabularnewline
9 & 10:54:31.70 & -21:04:44.5 & 17.01(0.01) & 16.39(0.01) & 16.12(0.01)    \tabularnewline
\hline
\end{tabular}
\tablefoot{Magnitudes of sequence stars are taken from SDSS9 and shown to two decimal places in the AB system. 1$\sigma$ uncertainties are given in parentheses.}
\end{table}

\begin{table}[h!]
\centering
\caption{Local sequence stars used to calibrate SN~2015H LSQ photometry.}\tabularnewline
\label{tab:15h--lsq-std}\tabularnewline
\begin{tabular}{lllll}\hline
\hline
No.  & RA & DEC &  $g$ & $r$  \tabularnewline
& &  &  (mag) & (mag)   \tabularnewline
\hline
\hline
10 & 10:54:54.07 & -21:03:06.2 & 18.60(0.01) & 17.27(0.01)  \tabularnewline
11 & 10:54:52.72 & -21:02:52.6 & 17.68(0.01) & 17.28(0.01)  \tabularnewline
12 & 10:54:47.70 & -21:02:43.7 & 17.93(0.01) & 17.45(0.01)  \tabularnewline
\phn6  & 10:54:45.04 & -21:01:21.9 & 15.76(0.01) & 15.25(0.01)  \tabularnewline
13 & 10:54:39.36 & -21:01:21.5 & 19.56(0.01) & 18.35(0.01)  \tabularnewline
14 & 10:54:36.85 & -21:01:26.5 & 18.70(0.01) & 17.64(0.01)  \tabularnewline
\hline
\end{tabular}
\tablefoot{Magnitudes of sequence stars are taken from SDSS9 and shown to two decimal places in the AB system. 1$\sigma$ uncertainties are given in parentheses.}
\end{table}

\begin{table}[h!]
\centering
\caption{Local sequence stars used to calibrate SN~2015H NTT photometry.}\tabularnewline
\label{tab:15h--ntt-std}\tabularnewline
\begin{tabular}{lllllll}\hline
\hline
No.  & RA & DEC &  $V$ & $J$ & $H$ & $K$  \tabularnewline
& & & (mag) & (mag) & (mag) & (mag)   \tabularnewline
\hline
\hline
\phn3  & 10:54:48.88 & -21:04:33.8 & 17.89(0.01)  &  14.77(0.03) & 14.07(0.04) & 13.91(0.05)   \tabularnewline
15 & 10:54:48.35 & -21:04:59.1 & 18.14(0.02)  &  14.58(0.04) & 13.93(0.04) & 13.69(0.04)   \tabularnewline
16 & 10:54:45.45 & -21:05:47.3 & 15.01(0.01)  &  13.41(0.03) & 12.92(0.03) & 12.92(0.03)   \tabularnewline
17 & 10:54:38.89 & -21:04:33.8 & 17.35(0.01)  &  16.20(0.08) & 15.80(0.14) & 15.87(0.28)   \tabularnewline
18 & 10:54:35.85 & -21:03:36.2 & 15.98(0.01)  &  14.70(0.04) & 14.34(0.04) & 14.20(0.07)   \tabularnewline
\hline
\end{tabular}
\tablefoot{$V$-- band magnitudes of sequence stars are derived from SDSS9 $gr$ magnitudes, and shown to two decimal places in the AB system. IR magnitudes are taken from 2MASS Point Source Catalogue in the 2MASS natural system. 1$\sigma$ uncertainties are given in parentheses.}
\end{table}

\begin{longtable}{llllllll}
\caption{Optical photometric journal for  SN~2015H}\tabularnewline
\label{tab:15h--phot--opt}\tabularnewline
\hline\hline
Date  & MJD & Phase & $g$ & $V$ & $r$ & $i$& Telescope/Survey\tabularnewline
 & & (days) & (mag)  & (mag) & (mag) & (mag) & \tabularnewline
\hline\hline
\endfirsthead
\caption{continued.}\tabularnewline
\hline\hline
Date  & MJD & Phase & $g$ & $V$ & $r$ & $i$& Telescope/Survey\tabularnewline
 & & (days) & (mag) & (mag) & (mag) & (mag) & \tabularnewline
\hline\hline
\endhead
\hline
\endfoot
2015 Jan.  23 &	57045.14	&  $-$17    & ...           & ...				& \ga20.2 &	...	&	LSQ \tabularnewline
2015 Jan.  25 &	57047.13 	&  $-$15  	& ... 			& ...	 			& 19.71(0.19) 	& ... 			& LSQ 	\tabularnewline
2015 Jan.  27 &	57049.12 	&  $-$13  	& ... 			& ...	 			& 18.55(0.10) 	& ... 			& LSQ 	\tabularnewline
2015 Jan.  29 &	57051.11 	&  $-$11  	& ... 			& ...	 			& 17.73(0.07) 	& ... 			& LSQ 	\tabularnewline
2015 Jan.  31 &	57053.11 	&  \phn$-$9  		& ... 			& ...	 			& 17.33(0.09) 	& ... 			& LSQ 	\tabularnewline
2015 Feb.  02 &	57055.10 	&  \phn$-$7  		& ... 			& ...	 			& 17.08(0.11) 	& ... 			& LSQ 	\tabularnewline
2015 Feb.  04 &	57057.10 	&  \phn$-$5  		& ... 			& ...	 			& 17.02(0.05) 	& ... 			& LSQ 	\tabularnewline
2015 Feb.  06 &	57059.09 	&  \phn$-$3  		& ... 			& ...	 			& 16.81(0.05) 	& ... 			& LSQ 	\tabularnewline
2015 Feb.  12 &	57065.08 	&  \,\phn\phn3     	& ... 			& ...	 			& 16.85(0.02) 	& ... 			& LSQ 	\tabularnewline
2015 Feb.  12 &	57065.13 	&  \,\phn\phn3  		& ... 			& 17.13(0.03)		& ... 			& ... 			& NTT 	\tabularnewline
2015 Feb.  14 &	57067.93 	&  \phn\phn6  		& 17.79(0.04)	& ...	 			& 16.99(0.02) 	& 16.98(0.02) 	& CPT 	\tabularnewline
2015 Feb.  15 &	57068.24 	&  \phn\phn6  		& ... 			& 17.40(0.02)		& ... 			& ... 			& NTT 	\tabularnewline
2015 Feb.  16 &	57069.81 	&  \phn\phn8  		& 18.06(0.03)	& ... 				& 17.05(0.02) 	& 17.00(0.04) 	& CPT 	\tabularnewline
2015 Feb.  17 &	57070.07 	&  \phn\phn8  		& ... 			& ...	 			& 17.04(0.02) 	& ... 			& LSQ 	\tabularnewline
2015 Feb.  19 &	57072.15 	&  \phn10  		& ... 			& ...	 			& 17.17(0.04) 	& ... 			& LSQ 	\tabularnewline
2015 Feb.  19 &	57072.24 	&  \phn10  		& ... 			& 17.75(0.02)		& ... 			& ... 			& NTT 	\tabularnewline
2015 Feb.  19 &	57072.59 	&  \phn11  		& 18.33(0.04)	& ... 				& 17.18(0.02) 	& 17.11(0.02) 	& COJ 	\tabularnewline
2015 Feb.  22 &	57075.32 	&  \phn13  		& 18.67(0.04)	& ... 				& 17.42(0.02) 	& 17.23(0.02) 	& LSC 	\tabularnewline
2015 Feb.  23 &	57076.06 	&  \phn14  		& ... 			& ...	 			& 17.37(0.02) 	& ... 			& LSQ 	\tabularnewline
2015 Feb.  25 &	57078.12 	&  \phn16  		& ... 			& 18.11(0.04)		& ... 			& ... 			& NTT 	\tabularnewline
2015 Feb.  25 & 57078.30 	&  \phn16  		& 18.91(0.03)	& ...	 			& 17.59(0.02)	& 17.43(0.03)	& LSC 	\tabularnewline
2015 Feb.  26 &	57079.10 	&  \phn17  		& ... 			& 18.10(0.02)		& ... 			& ... 			& NTT 	\tabularnewline
2015 Feb.  26 &	57079.80 	&  \phn18  		& 18.80(0.07)	& ... 				& 17.55(0.05) 	& 17.30(0.07) 	& CPT 	\tabularnewline
2015 Mar.  01 &	57082.05 	&  \phn20  		& ... 			& ...	 			& 17.81(0.04) 	& ... 			& LSQ 	\tabularnewline
2015 Mar.  01 &	57082.81 	&  \phn21  		& 19.09(0.07)	& ... 				& 17.95(0.04) 	& 17.47(0.07)	& CPT 	\tabularnewline
2015 Mar.  03 &	57084.07 	&  \phn22  		& ... 			& ...	 			& 17.87(0.05) 	& ... 			& LSQ 	\tabularnewline
2015 Mar.  04 &	57085.51 	&  \phn24  		& 19.19(0.07)	& ... 				& 18.00(0.02) 	& 17.61(0.06) 	& COJ 	\tabularnewline
2015 Mar.  05 &	57086.05 	&  \phn24  		& ... 			& ...	 			& 17.97(0.03) 	& ... 			& LSQ 	\tabularnewline
2015 Mar.  07 &	57088.05 	&  \phn26  		& ... 			& ...	 			& 17.99(0.05) 	& ... 			& LSQ 	\tabularnewline
2015 Mar.  07 &	57088.22 	&  \phn26  		& 19.28(0.05)	& ... 				& 18.10(0.03) 	& 17.76(0.04) 	& LSC 	\tabularnewline
2015 Mar.  08 &	57089.92 	&  \phn28  		& 19.30(0.08)	& ... 				& 18.04(0.05)	& 18.05(0.07) 	& CPT 	\tabularnewline
2015 Mar.  09 &	57090.08 	&  \phn28  		& ... 			& ...	 			& 18.13(0.05) 	& ... 			& LSQ 	\tabularnewline
2015 Mar.  10 &	57091.63 	&  \phn30  		& 19.27(0.10) 	& ... 				& 18.15(0.05) 	& 17.97(0.09) 	& COJ 	\tabularnewline
2015 Mar.  11 &	57092.07 	&  \phn30  		& ... 			& ...	 			& 18.23(0.04) 	& ... 			& LSQ 	\tabularnewline
2015 Mar.  12 &	57093.05 	&  \phn31  		& ... 			& 18.71(0.04)		& ... 			& ... 			& NTT 	\tabularnewline
2015 Mar.  13 &	57094.04 	&  \phn32  		& ... 			& ...	 			& 18.22(0.03) 	& ... 			& LSQ 	\tabularnewline
2015 Mar.  14 &	57095.05 	&  \phn33  		& ... 			& ...	 			& 18.27(0.05) 	& ... 			& LSQ 	\tabularnewline
2015 Mar.  16 &	57097.05 	&  \phn35  		& ... 			& ...	 			& 18.32(0.05) 	& ... 			& LSQ 	\tabularnewline
2015 Mar.  16 &	57097.47 	&  \phn36  		& ... 			& ... 				& 18.28(0.07) 	& 18.03(0.13) 	& COJ 	\tabularnewline
2015 Mar.  19 &	57100.20 	&  \phn38  		& ... 			& ... 				& 18.42(0.06) 	& 18.27(0.06) 	& LSC 	\tabularnewline
2015 Mar.  20 &	57101.05 	&  \phn39  		& ... 			& ...	 			& 18.39(0.04) 	& ... 			& LSQ 	\tabularnewline
2015 Mar.  21 &	57102.90 	&  \phn41  		& ... 			& ... 				& 18.37(0.06) 	& 18.39(0.10) 	& CPT 	\tabularnewline
2015 Mar.  22 &	57103.04 	&  \phn41  		& ... 			& ...	 			& 18.47(0.06) 	& ... 			& LSQ 	\tabularnewline
2015 Mar.  24 &	57105.73 	&  \phn44  		& ... 			& ... 				& 18.40(0.06) 	& 18.27(0.09) 	& CPT 	\tabularnewline
2015 Mar.  27 &	57108.45 	&  \phn47  		& ... 			& ... 				& 18.46(0.05) 	& 18.32(0.09) 	& COJ 	\tabularnewline
2015 Mar.  28 &	57109.15 	&  \phn47  		& ... 			& 18.99(0.03) 		& ...	 		& ...	 		& NTT 	\tabularnewline
2015 Mar.  30 &	57111.16 	&  \phn49  		& ... 			& ... 				& 18.70(0.05) 	& 18.40(0.08) 	& LSC 	\tabularnewline
2015 Apr.  08 &	57120.72 	&  \phn59  		& ... 			& ... 				& ... 			& 18.53(0.07) 	& CPT 	\tabularnewline
2015 Apr.  11 &	57123.43 	&  \phn62  		& ... 			& ... 				& 18.59(0.07) 	& 18.52(0.07) 	& COJ 	\tabularnewline
2015 Apr.  14 &	57126.14 	&  \phn64  		& ... 			& ... 				& 18.96(0.05) 	& 18.60(0.08) 	& LSC 	\tabularnewline
2015 Apr.  16 &	57128.89 	&  \phn67  		& ... 			& ... 				& 18.97(0.15) 	& 18.75(0.25) 	& CPT 	\tabularnewline
2015 Apr.  19 &	57131.71 	&  \phn70  		& ... 			& ... 				& 19.05(0.08) 	& 18.86(0.11) 	& CPT 	\tabularnewline
2015 Apr.  22 &	57134.71 	&  \phn73  		& ... 			& ... 				& 19.02(0.10) 	& 18.69(0.11) 	& CPT 	\tabularnewline
2015 Apr.  25 &	57137.97 	&  \phn76  		& ... 			& ... 				& 19.17(0.09) 	& 18.94(0.08) 	& CPT 	\tabularnewline
2015 Apr.  30 &	57142.76 	&  \phn81  		& ... 			& ... 				& 18.95(0.13) 	& 18.66(0.12) 	& CPT 	\tabularnewline
2015 May  05 &	57147.40 	&  \phn86  		& ... 			& ... 				& 18.98(0.12) 	& 19.02(0.15) 	& COJ 	\tabularnewline
2015 May  10 &	57152.71 	&  \phn91  		& ... 			& ... 				& 19.25(0.16) 	& 18.93(0.06) 	& CPT 	\tabularnewline
2015 May  13 &	57155.44 	&  \phn94  		& ... 			& ... 				& 19.20(0.05) 	& 19.01(0.04) 	& COJ 	\tabularnewline
2015 May  18 &	57160.43 	&  \phn99  		& ... 			& ... 				& 19.19(0.10) 	& 18.99(0.07) 	& COJ 	\tabularnewline
2015 May  23 &	57165.99 	&  104   	& ... 			& ... 				& 19.58(0.09) 	& 19.31(0.09) 	& LSC 	\tabularnewline
2015 May  28 &	57170.78 	&  109   	& ... 			& ... 				& 19.39(0.07) 	& 19.35(0.16) 	& CPT 	\tabularnewline
2015 Jun.  02 &	57175.99 	&  114   	& ... 			& ... 				& 19.40(0.06) 	& 19.27(0.09) 	& LSC 	\tabularnewline
2015 Jun.  03 &	57176.40 	&  115   	& ... 			& ... 				& 19.50(0.05) 	& 19.32(0.07) 	& COJ 	\tabularnewline
2015 Jun.  09 &	57182.39 	&  120   	& ... 			& ... 				& 19.56(0.06) 	& 19.91(0.21) 	& COJ 	\tabularnewline
2015 Jun.  14 &	57187.37 	&  125   	& ... 			& ... 				& 19.80(0.12) 	& 19.45(0.11) 	& COJ 	\tabularnewline
2015 Jun.  19 &	57192.73 	&  131   	& ... 			& ... 				& 19.84(0.16) 	& 19.44(0.17) 	& CPT 	\tabularnewline
2015 Jun.  28 &	57201.69 	&  140  	& ... 			& ... 				& 19.70(0.13) 	& 19.66(0.21) 	& CPT 	\tabularnewline
2015 Jul.  04 &	57207.72 	&  146  	& ... 			& ... 				& 19.80(0.22) 	& 19.55(0.20) 	& CPT 	\tabularnewline\hline

\end{longtable}
\tablefoot{1$\sigma$ uncertainties are given in parentheses. CPT, COJ, and LSC refer to telescopes that are part of the LCOGT 1-m network. CPT is located at the South African Astronomical Observatory (SAAO), South Africa; COJ is located at the Siding Spring Observatory (SSO), Australia; LSC is located at the Cerro Tololo Inter-American Observatory (CTIO), Chile. 
}

\begin{table}[h!]
\centering
\caption{IR photometric journal for  SN~2015H}\tabularnewline
\label{tab:15h--phot--ir}\tabularnewline
\begin{tabular}{lllllll}
\hline\hline
Date & MJD & Phase & $J$ &  $H$ &  $K$ & Telescope\tabularnewline
& & (days) & (mag) &  (mag) &  (mag) & \tabularnewline
\hline\hline
2015 Feb. 16 &	57069.29 	& \phn7	& 16.85(0.03)	& 16.49(0.04) 	& 16.37(0.06) 	& NTT 	\tabularnewline
2015 Mar. 12 &	57093.05 	& 31	& 17.56(0.03) 	& 17.11(0.03) 	& ... 			& NTT 	\tabularnewline
2015 Mar. 29 &	57110.23 	& 48	& 18.11(0.04) 	& 17.54(0.05) 	& ... 			& NTT 	\tabularnewline
\hline
\end{tabular}
\tablefoot{1$\sigma$ uncertainties are given in parentheses.}
\end{table}

\begin{table}[h!]
\centering
\caption{Spectroscopic journal for  SN~2015H}\tabularnewline
\label{tab:15h--specseq}\tabularnewline
\begin{tabular}{lllllll}\hline
\hline
Date  & MJD & Phase &  Instrument & Grism & Wavelength Coverage & Resolution \tabularnewline
  &  & (days) &   &  &  (\AA) &  (\AA) \tabularnewline

\hline
\hline
2015 Feb. 11 &       	57065.13 & \phn\phn3  	& EFOSC2  	& Gr \#13 			 & 3650 -- 9250 & 17.5  \tabularnewline 
2015 Feb. 14 &	   		57068.24 & \phn\phn6  	& EFOSC2 	& Gr \#11 \& Gr \#16 & 3345 -- 9995 & 14.2 \& 13.8  \tabularnewline
2015 Feb. 15 &   		57069.18 & \phn\phn7  	& SOFI 		& Blue \& Red 		 & 9350 -- 25\,360 & 23.7 \& 28.8  \tabularnewline
2015 Feb. 18 &       	57072.28 & \phn10  	& EFOSC2 	& Gr \#11 \& Gr \#16 & 3345 -- 9995 & 14.1 \& 12.5  \tabularnewline
2015 Feb. 24 &      	57078.12 & \phn16  	& EFOSC2 	& Gr \#11 			 & 3345 -- 7470 & 21.0  \tabularnewline 
2015 Feb. 24 &	  		57078.20 & \phn16  	& SOFI 		& Blue  			& 9350 -- 16\,450 & 22.2  \tabularnewline
2015 Feb. 25 &       	57079.10 & \phn17  	& EFOSC2 	& Gr \#16 			& 6000 -- 9995 & 20.7  \tabularnewline
2015 Mar. 10 &      	57092.29 & \phn30  	& SOFI 		& Blue  			& 9350 -- 16\,450 & 24.1  \tabularnewline
2015 Mar. 11 &       	57093.05 & \phn31  	& EFOSC2 	& Gr \#13 			& 3650 -- 9250 & 17.8  \tabularnewline
2015 Mar. 27 &      	57109.15 & \phn47  	& EFOSC2 	& Gr \#13 			& 3650 -- 9250 & 17.4  \tabularnewline 
2015 Jun. 01 &      	57174.88 & 113  & OSIRIS 	& R500R 			& 4800 -- 10\,000 & 15.6  \tabularnewline 
\hline
\end{tabular}
\tablefoot{
Phases are given relative to an estimated $r$--band maximum of MJD = 57061.9. Resolutions are measured from the FWHM of sky lines. Note that the slit width for EFOSC2 exposures taken on 2015 Feb 24 and 2015 Feb 25 was increased to 1.5". All other exposures were obtained with 1$"$ slit widths.
}
\end{table}

\begin{table}[h!]
\centering
\caption{Optical photometric journal for  PS15csd}\tabularnewline
\label{tab:15csd--phot--opt}\tabularnewline
\begin{tabular}{llllll}
\hline\hline
Date & MJD & Phase & $g$ & $r$ & $i$ \tabularnewline
 & & (days) & (mag)  & (mag) & (mag) \tabularnewline
\hline\hline
2015 Nov. 11 &	57338.04 	&  \phn1  	& 19.37(0.03)			& 18.60(0.02)	 			& 18.72(0.02) 	\tabularnewline
2015 Nov. 16 &	57343.00 	&  \phn6  	& 20.22(0.08) 			& 18.93(0.02)	 			& 18.80(0.04) 	\tabularnewline
2015 Nov. 17 &	57343.93 	&  \phn7  	& 20.20(0.05) 			& 18.98(0.02)	 			& 18.88(0.03) 	\tabularnewline
2015 Nov. 19 &	57345.98 	&  \phn9  	& 20.69(0.05)			& 19.19(0.04)	 			& 18.89(0.03) 	\tabularnewline
2015 Nov. 21 &	57347.86 	&  11  	& 20.81(0.07) 			& 19.42(0.03)	 			& 19.03(0.03) 	\tabularnewline
2015 Nov. 26 &	57352.94 	&  16  	& 21.59(0.19) 			& 19.64(0.07)	 			& 19.29(0.07) 	\tabularnewline
2015 Dec. 03 &	57359.93 	&  23  	& 21.45(0.16) 			& 19.80(0.06)	 			& 19.54(0.06) 	\tabularnewline
\hline
\end{tabular}
\tablefoot{1$\sigma$ uncertainties are given in parentheses. All observations were obtained via the Liverpool Telescope. Phases are given relative to an estimated date of $r$--band maximum MJD = 57337.3. }
\end{table}
\twocolumn

\begin{table}[h!]
\centering
\caption{References for comparison SNe used throughout this paper.}\tabularnewline
\label{tab:objs}\tabularnewline
\begin{tabular}{lll}\hline
\hline
SN  & SN Type & Reference \tabularnewline
\hline
\hline
2002cx		& Iax	&	1	\tabularnewline
2003gq		& Iax	&	1	\tabularnewline
2004cs		& Ia	&	1	\tabularnewline
2004eo		& Ia	&	2	\tabularnewline
2005cc		& Iax	&	1	\tabularnewline
2005cf		& Ia	&	3	\tabularnewline
2005hk		& Iax 	&	4	\tabularnewline
2007qd		& Iax	&	5	\tabularnewline
2008A		& Iax	&	1	\tabularnewline
2008ae		& Iax	&	1	\tabularnewline
2008ha		& Iax	&	1	\tabularnewline
2009J		& Iax	&	1	\tabularnewline
2009ku		& Iax	&	1	\tabularnewline
2010ae		& Iax	&	6	\tabularnewline
2011ay		& Iax	&	1	\tabularnewline
2011fe		& Ia	&	7	\tabularnewline
2012Z		& Iax	&	4	\tabularnewline
PS15csd		& Iax	&	8	\tabularnewline
PTF09ego	& Iax	&	9	\tabularnewline
PTF09eoi	& Iax	&	9	\tabularnewline
PTF11hyh	& Iax	&	9	\tabularnewline
\hline
\end{tabular}
\tablefoot{(1) \cite{foley--13}; (2) \cite{04eo}; (3) \cite{05cf}; (4) \cite{comp--obs--12z}; (5) \cite{obs--07qd}; (6) \cite{obs--10ae}; (7) \cite{11fe--nature}; (8) this work; (9) \cite{slow--ptf}}
\end{table}

\twocolumn
\section{Example TARDIS model for SN~2015H}
\label{apdx:tardis}

Here we describe the TARDIS model used to generate a synthetic spectrum of SN~2015H, as shown in Fig.~\ref{fig:15h--tardis-id}. We stress that the model presented here provides a good match to the observed spectrum, but is not a unique solution given the inevitable degeneracy between the input parameters.
\par
TARDIS is a 1D Monte Carlo radiative transfer code that assumes the SN ejecta are spherically symmetric. The model domain is divided into a number of equally spaced concentric zones, defined by an inner and an outer boundary. A TARDIS calculation consists of an iterative sequence of Monte Carlo radiative transfer simulations. Each Monte Carlo packet, representing a bundle of identical photons, has frequency randomly selected from a blackbody distribution and is injected into the model at the inner boundary and propagated through the model region. Once all packet propagation is complete, the packet flight histories are used to estimate local ionization/excitation conditions in each zone, and to compute an emergent spectrum. The escaping packets are also used to make an improved estimate of the input blackbody temperature required to match a user-specified value for the total emergent luminosity. The new ionization/excitation conditions and blackbody temperature are adopted in the next iteration, duly leading to a consistent model. The final emergent spectrum can then be analysed and compared to the observations. 
 
\par
In the model adopted here, we use a uniform composition and exponential density profile for the SN ejecta

\begin{equation} 
\label{eq:dens-pro}
\centering
\rho(v, t_\mathrm{{exp}}) = \rho_\mathrm{0} \; \left( { \frac{t_\mathrm{0}}{t_\mathrm{{exp}}} } \right)^3 \exp(-v/v_\mathrm{0})
\end{equation}
where $v$ is the velocity, $t_\mathrm{{exp}}$ is the time since explosion, and $v_\mathrm{0}$ and $t_\mathrm{0}$ are a reference velocity and time, respectively. The boundaries of the model are set in velocity space at values guided by the velocities measured in SN~2015H as shown in Fig.~\ref{fig:15h--velocities}. Values for the emergent luminosity and $t_\mathrm{{exp}}$ are chosen based on our photometric measurements (see Section~\ref{sect:photometric} for discussion on the explosion date of SN~2015H).
\par

\begin{table}
\centering
\caption{Input parameters used in modelling SN~2015H spectrum with TARDIS}\tabularnewline
\label{tab:15h--tardis}\tabularnewline
\begin{tabular}{ll}\hline
\hline
Supernova properties&  \tabularnewline
\hline
Emergent Luminosity 							&	8.53 log L$_{\odot}$ 		\tabularnewline
Time since explosion, $t_\mathrm{{exp}}$		& 	22 days 					\tabularnewline
\hline
Density profile &	\tabularnewline
\hline
Reference time, $t_\mathrm{0}$				&	2 days						\tabularnewline
Reference density, $\rho_\mathrm{0}$			&	2.2 $\times$ 10$^{-10}$ g~cm$^{-3}$	\tabularnewline
Reference velocity, $v_\mathrm{0}$			&	3000 km~s$^{-1}$			\tabularnewline
\hline
Velocity boundaries &	\tabularnewline
\hline
Inner			&	5500 km~s$^{-1}$			\tabularnewline
Outer			&	7500 km~s$^{-1}$			\tabularnewline
\hline
\end{tabular}
\end{table}

In Table~\ref{tab:15h--tardis-comp} we compare the composition of the ejecta in our TARDIS model to the deflagration models N3def \citep{3d--deflag--sim--obs} and 2D70 \citep{kicked--remnants}. We developed the model ejecta composition by initially including only \ion{Ca}{} and fitting the \ion{Ca}{ii} IR triplet profile. Having obtained good agreement, we progressively added the elements expected in SNe Iax spectra guided by explosive nucleosynthesis models \citep{3d--deflag--sim--obs}. 
Some elements (such as Ne, Al, etc.) were found to have a negligible effect on our model spectrum we therefore do not include them in the TARDIS model, despite them having significant abundances in explosive nucleosynthesis models. In such cases the abundances should be regarded as unconstrained by our TARDIS model.
\par
We find that our preferred ejecta composition is broadly consistent with that of the N3def model published by \cite{3d--deflag--sim--obs}.  However, there are some notable differences: in particular, compared to the N3def \citep{3d--deflag--sim--obs} and 2D70 \citep{kicked--remnants} models, our empirically derived composition favours a lower mass fraction of IMEs.
This may be due to the relatively late epochs of our observed spectra, where any strong signatures of IMEs are masked by IGEs. Earlier spectra may have provided more robust constraints on the abundance of IMEs, and/or any potential stratification in the ejecta.

\begin{table}
\centering
\caption{Comparison of ejecta compositions for SNe Iax deflagration models 
}\tabularnewline
\label{tab:15h--tardis-comp}\tabularnewline
\begin{tabular}{llll}\hline
\hline
Abundances &	\tabularnewline
\hline
& TARDIS & N3def & 2D70 \tabularnewline
\hline
C-O:	& 0.302	&	0.305(0.369)	 & 0.565	\tabularnewline
IMEs:	& 0.043 & 	0.132(0.087)	 & 0.130	\tabularnewline
IGEs:	& 0.655	&	0.543(0.526)	 & 0.304	\tabularnewline
\hline
Helium		&								&	$4.00(4.05) \times  10^{-4}$				\tabularnewline
Carbon		&	$1.29 \times 10^{-1}$		&	$1.22(1.61) \times  10^{-1}$				\tabularnewline	
Nitrogen	&								&	$6.26(5.10) \times  10^{-6}$				\tabularnewline
Oxygen		&	$1.72 \times 10^{-1}$		&	$1.83(2.08) \times  10^{-1}$				\tabularnewline	
Fluorine	&								&	$9.77(7.03) \times  10^{-9}$				\tabularnewline
Neon		&								&	$1.82(1.66) \times  10^{-2}$				\tabularnewline
Sodium		&	$8.62 \times 10^{-3}$		&	$2.08(1.55) \times  10^{-4}$				\tabularnewline	
Magnesium	&	$8.62 \times 10^{-3}$		&	$1.59(1.09) \times  10^{-2}$				\tabularnewline	
Aluminium	&								&	$1.05(0.70) \times  10^{-3}$				\tabularnewline
Silicon		&	$8.62 \times 10^{-3}$		&	$8.01(5.25) \times  10^{-2}$				\tabularnewline	
Phosphorus	&								&	$3.98(2.72) \times  10^{-4}$				\tabularnewline
Sulfur		&	$8.62 \times 10^{-3}$		&	$2.69(1.78) \times  10^{-2}$				\tabularnewline
Chlorine	&								&	$7.67(5.21) \times  10^{-5}$				\tabularnewline
Argon		&								&	$4.28(2.88) \times  10^{-3}$				\tabularnewline
Potassium	&								&	$2.67(1.67) \times  10^{-5}$				\tabularnewline
Calcium		&	$8.62 \times 10^{-3}$		&	$3.15(2.19) \times  10^{-3}$				\tabularnewline	
Scandium	&								&	$1.08(0.89) \times  10^{-8}$				\tabularnewline
Titanium	&								&	$5.72(4.25) \times  10^{-5}$				\tabularnewline
Vanadium	&								&	$3.15(2.41) \times  10^{-5}$				\tabularnewline
Chromium	&	$8.62 \times 10^{-3}$		&	$3.28(2.97) \times  10^{-3}$				\tabularnewline	
Manganese	&								&	$2.03(2.28) \times  10^{-4}$				\tabularnewline
Iron		&	$2.59 \times 10^{-1}$		&	$8.16(9.05) \times  10^{-2}$				\tabularnewline	
Cobalt		&	$3.02 \times 10^{-1}$		&	$3.82(3.53) \times  10^{-1}$				\tabularnewline	
Nickel		&	$8.62 \times 10^{-2}$		&	$7.62(7.91) \times  10^{-2}$				\tabularnewline
Copper		&								&	$9.62(8.44) \times  10^{-5}$				\tabularnewline
Zinc		&								&	$6.14(5.85) \times  10^{-4}$				\tabularnewline

\hline
\end{tabular}
\tablefoot{N3def values are taken from the N3def model published by \cite{3d--deflag--sim--obs}. 2D70 values are taken from \cite{kicked--remnants}. Values given here represent mass fractions of each model. Values given in parentheses are mass fractions of the N3def model in the region with velocities between 5500 km~s$^{-1}$ and 7500 km~s$^{-1}$, the range of velocities covered in our TARDIS calculations.  Here, we define IMEs as elements with atomic number between $Z=11$ and $Z=20$ (sodium to calcium), while IGEs refers to elements with $Z=21$ to 28 (scandium to nickel). Elements not included in the TARDIS model were found to have negligible impact on the spectrum, and are therefore unconstrained by our modelling. 
}
\end{table}

\end{appendix}

\end{document}